\documentclass[graybox]{svmult}
\usepackage{mathptmx}      
\usepackage{helvet}         
\usepackage{courier}       
\usepackage{type1cm}      
\usepackage{makeidx}       
\usepackage{graphicx}       
\usepackage{multicol}      
\usepackage[bottom]{footmisc}
\usepackage{amsmath}

\makeindex            

\begin{document}

\title*{Symmetry and topology in antiferromagnetic spintronics}
\author{Libor \v{S}mejkal and Tom\'a\v{s} Jungwirth}
\institute{Libor \v{S}mejkal \at Institut f\"ur Physik, Johannes Gutenberg-Universit\"at, Staudinger Weg 7, 55128 Mainz, Germany, Institute of Physics, Academy of Sciences of the Czech Republic,
Cukrovarnicka 10, 162 00 Praha 6, Czech Republic, Faculty of Mathematics and Physics, Charles University,
Department of Condensed Matter Physics, Ke Karlovu 5, 12116 Praha 2, Czech Republic, \email{smejkall@fzu.cz}
\and  Tom\'a\v{s} Jungwirth \at  Institute of Physics, Academy of Sciences of the Czech Republic,
Cukrovarnicka 10, 162 00 Praha 6, Czech Republic, School of Physics and Astronomy, University of Nottingham,
University Park, Nottingham NG7 2RD, United Kingdom, \email{jungw@fzu.cz}}

\maketitle

\abstract{Antiferromagnetic spintronics focuses on investigating and using antiferromagnets as active elements in spintronics structures. Last decade advances in relativistic spintronics led to the discovery of the staggered, current-induced field in antiferromagnets. The corresponding N\'eel spin-orbit torque allowed for efficient electrical switching of antiferromagnetic moments and, in combination with electrical readout, for the demonstration of experimental antiferromagnetic memory devices. In parallel, the anomalous Hall effect was predicted and subsequently observed in antiferromagnets. A new field of spintronics based on antiferromagnets has emerged. We will focus here on the introduction into the most significant discoveries which shaped the field together with a more recent spin-off focusing on combining antiferromagnetic spintronics with topological effects, such as antiferromagnetic topological semimetals and insulators, and the interplay of antiferromagnetism, topology, and superconductivity in heterostructures.}

\section{Introduction}
\label{sec:1}
The phase of matter can be characterized by symmetry and topology \cite{Hasan2010,Haldane2017}. For certain effects, symmetry provides the basic condition to occur while topology can add exceptional robustness. A text-book example is the Hall effect enabled by the broken time-reversal symmetry ($\cal{T}$) in the applied magnetic field. The exact discrete resistance values in the quantum Hall effect (QHE), unperturbed by disorder, are then a consequence of the topological Landau level form of the electronic structure in a strong quantizing magnetic field \cite{Hansson2017}. In this chapter, we show how the fundamental concepts of symmetry and topology apply to antiferromagnetic spintronics \cite{Jungwirth2016}. 

We start our chapter by briefly illustrating the symmetry and topology principles on three key functionalities of spintronic memory devices, namely the retention, reading, and writing of magnetic information \cite{Chappert2007,Kent2015}. Side by side we compare in this introductory section how the principles apply when considering the more conventional ferromagnetic and the emerging antiferromagnetic spintronic devices.

Ferromagnetism can (and often does) lower the symmetry of the crystal, depending on the direction of magnetic moments. For example, a rotation along a certain crystal axis remains a symmetry operation when moments are aligned with the axis but the symmetry is broken when the moments are perpendicular to the rotation axis. It implies that reorientation of ferromagnetic moments can, in the presence of spin-orbit coupling (SOC), change the electronic structure and by this the total energy. This is the origin of the magnetocrystalline anisotropy energy (MAE) barrier that supports the non-volatile storage in spintronic memories \cite{Chappert2007}. The same symmetry principle and corresponding magnetic storage functionality apply equally to antiferromagnets (AFs) \cite{Jungwirth2016}. On top of that, the lack of a net magnetic moment and suppressed dipolar fields make the storage in AFs less sensitive to magnetic fields and allow for denser integration of memory bits than in ferromagnets. Apart from binary storage, AFs naturally host series of different stable multi-domain reconfigurations which is suitable for integrated memory-logic or neuromorphic computation devices \cite{Wadley2016,Kriegner2016,Olejnik2017,Borders2017}.

\begin{figure}[h]
\sidecaption
% For example, with the graphicx style use
\includegraphics[width=0.9\linewidth]{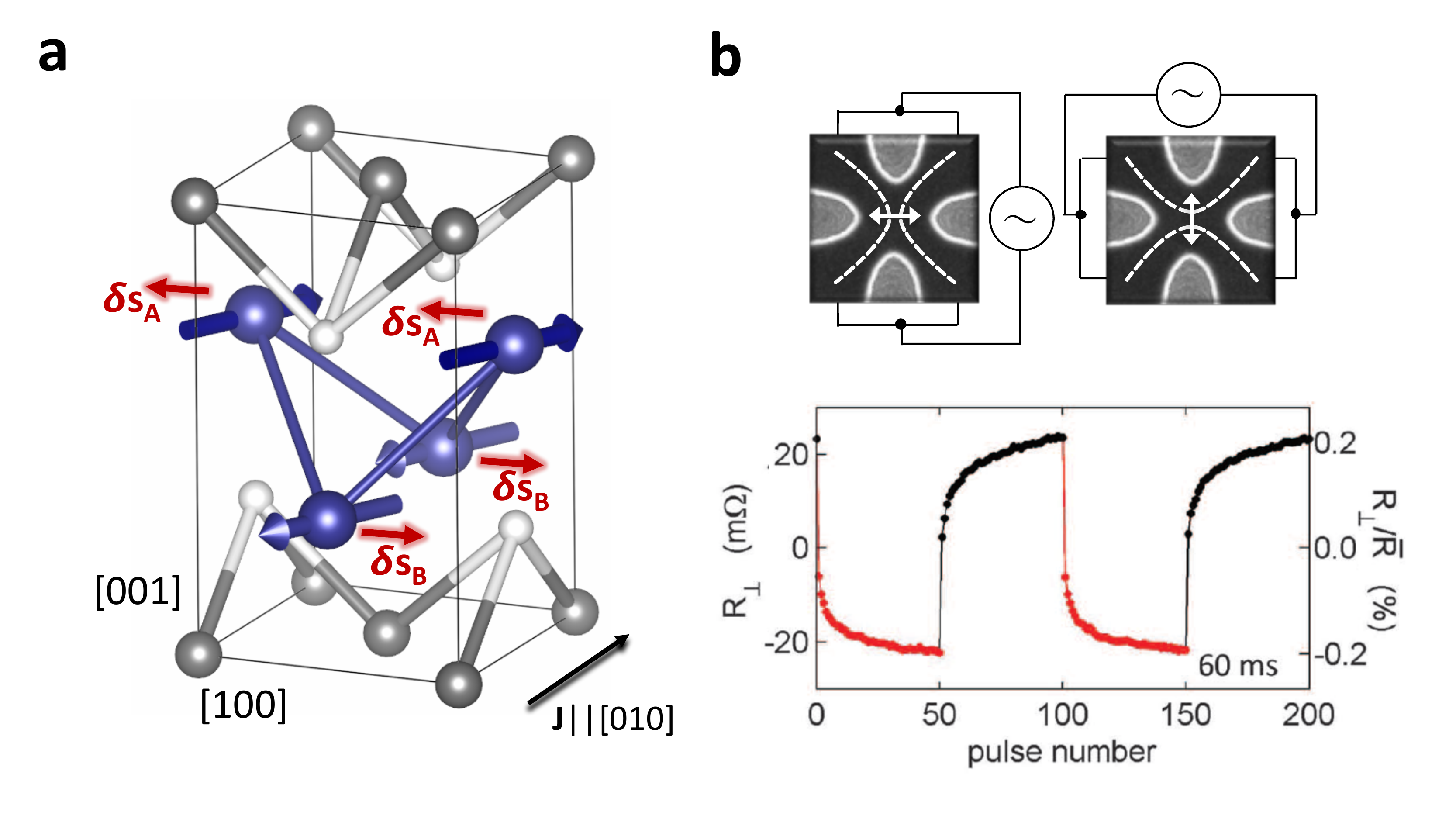}
%
% If no graphics program available, insert a blank space i.e. use
%\picplace{5cm}{2cm} % Give the correct figure height and width in cm
%
\caption{\textbf{Discovery of the manipulation of the antiferromagnetic order via electric currents.} (a) The crystal structure of the CuMnAs AF with marked non-equilibrium spin polarisations $\delta\textbf{s}$ generated by the current $\textbf{J}$ applied along the [010] direction. (b)  The reorientation of the moments is observed in the microbars of the CuMnAs. The writing current is applied along one of two orthogonal directions and generates a spin-orbit field which reorients the moments along the perpendicular direction to the applied current. In the bottom panel, we show the anisotropic magnetoresistance signal corresponding to the reorientation of the N\'{e}el order parameter. Panel (b) adapted from Ref. \cite{Olejnik2017a}, and Ref. \cite{Wadley2016}.}
\label{fig:1}       % Give a unique label
\end{figure}

Anisotropic resistance, i.e. the sensitivity of electronic transport to the current direction, requires broken cubic symmetry. Ferromagnetism where spins align with a specific crystal axis always breaks cubic symmetry. This implies that ferromagnets can have the anisotropic resistance. Here typically the leading dependence of the resistance on current direction is when measured with respect to the magnetization axis. The effect called (spontaneous) anisotropic magnetoresistance (AMR) is known for more than 150 years \cite{Thomson1856} and provides arguably the most straightforward means for electrically detecting different directions of ferromagnetic moments. AMR was used, e.g., in the first generation of magnetoresistive field sensors for hard-disk read-heads or for electrical readout in the first generation of magnetic random access memories (MRAMs) \cite{Daughton1992}. The same symmetry argument applies to the AMR in AFs where it has been used to demonstrate electrical readout in experimental memory devices, as shown in Fig. \ref{fig:1} \cite{Park2011b,Marti2014,Wadley2016,Kriegner2016,Olejnik2017}. On the other hand, AMR in AFs is not suitable for external magnetic field sensing because of the lack of the net magnetic moment. 
\begin{figure}[h]
\sidecaption
\includegraphics[width=0.9\linewidth]{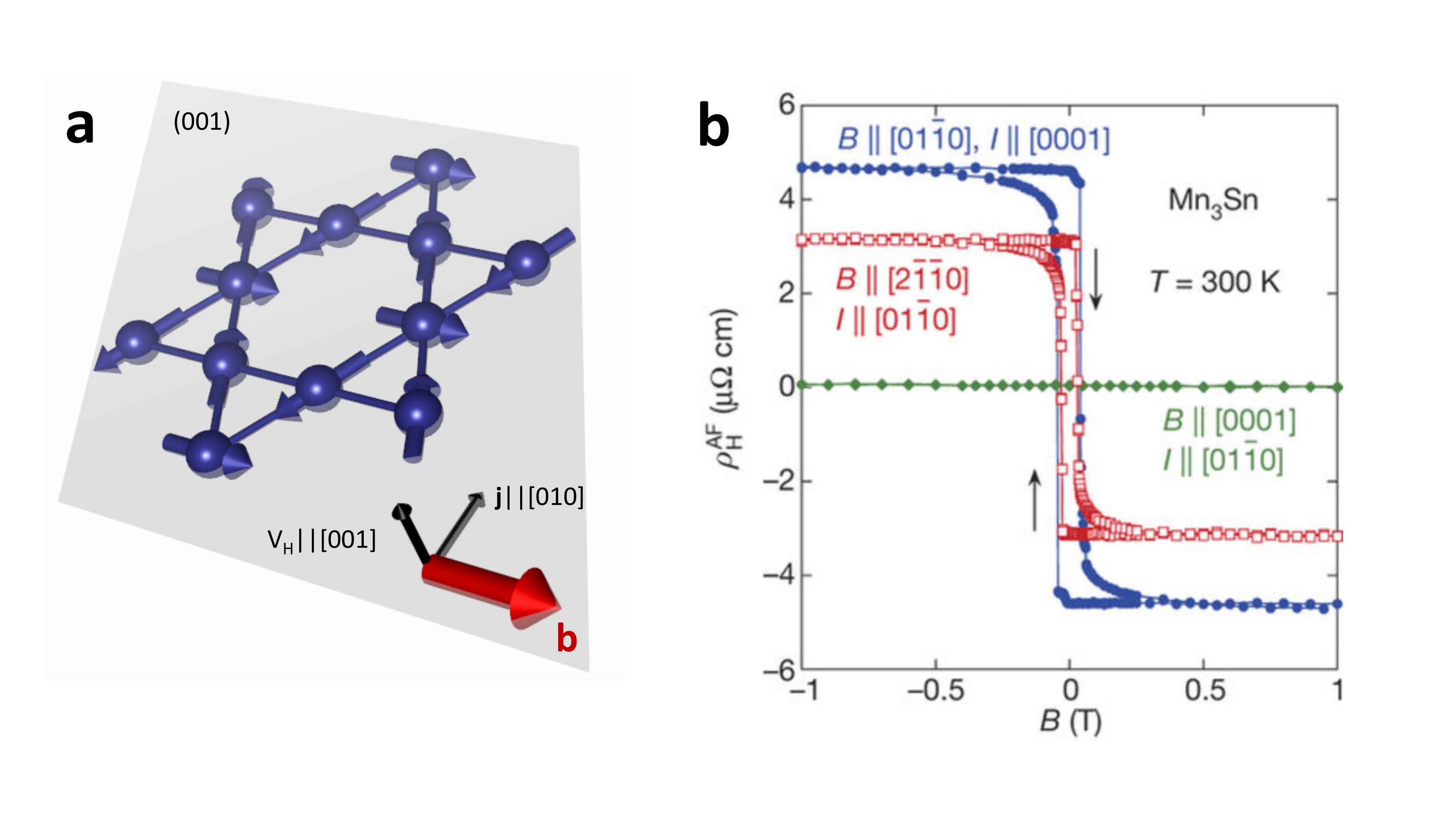}
\caption{\textbf{Observation of the anomalous Hall effect in noncollinear AFs.} (a) The noncollinear magnetic order on the kagome lattice in the Mn$_{\text{3}}$Sn breaks the time reversal symmetry and allows for the net magnetization in the [100] direction. Although the net moment is almost perfectly compensated, a strong emergent magnetic field, e.g. Berry curvature $\textbf{b}$, along the [100] direction generates a large anomalous Hall effect in the perpendicular, (001), plane. (b) The measured Hall resistivity $\rho_{H}^{AF}$ obtained by subtracting the signal from the small net moment and ordinary Hall effect. The panel (b) adapted from Ref. \cite{Nakatsuji2015}.}
\label{fig:2}       
\end{figure}

Another transport effect that can be used to internally detect magnetic moments of a conductor is the anomalous Hall effect (AHE). It can occur in crystals with magnetic space groups whose symmetry allows for the presence of a net magnetic moment. Remarkably, this symmetry argument holds independently of whether the system indeed has a ferromagnetic moment or is in a fully compensated antiferromagnetic state. While the AHE in ferromagnets was discovered more than 100 years ago \cite{Hall1881}, its experimental demonstration in AFs, shown in Fig. \ref{fig:2}, is one of the most recent developments in spintronics \cite{Nakatsuji2015}.

Unlike time-reversal, the symmetry operation of space-inversion ($\cal{P}$)  does not rotate the axial magnetic moment vector which implies that ferromagnets cannot have a combined $\cal{PT}$-symmetry. This removes the Kramers $\cal{PT}$-symmetry protection of the spin-up/spin-down degeneracy of electronic bands. As a result, electrons moving in the unequal spin-up and spin-down bands have different resistivities. In ferromagnetic bilayers this leads to different resistance states for parallel and antiparallel alignments of moments in the two layers and the corresponding giant/tunnelling magnetoresistance (GMR/TMR) effects \cite{Chappert2007}.  These phenomena, that tend be stronger than the AMR, are used in modern hard disk read-heads and MRAMs. 

Different resistivities in spin-up and spin-down transport channels in a ferromagnet can be also used to filter an unpolarised current passing through the ferromagnetic layer by suppressing one spin-component of the electrical current. The resulting spin-polarized current filtered through such a ferromagnetic polarizer can exert a spin transfer torque (STT) on the adjacent ferromagnetic layer and switch its magnetic moment \cite{Ralph2008}. This reversible electrical writing method is used in the latest generation of MRAMs. 

There is no equivalent counterpart of the two-spin-channel GMR/TMR or STT phenomena in AFs with equal spin-up and spin-down bands. On the other hand, AFs can have the combined $\cal{PT}$-symmetry which opens a possibility of the Dirac crossing of two doubly-degenerate bands, as shown in Fig. 3(a) \cite{Tang2016,Smejkal2016}. The topological protection of these Dirac points can be turned on and off by changing other symmetries of the antiferromagnetic crystal, e.g., via changing the direction of the antiferromagnetic N\'eel vector. This could enable very large topological AMR effects in AFs and remedy the absence of the two-spin-channel GMR/TMR.

In time-reversal symmetric paramagnets, a broken space-inversion symmetry leads to the Kramers degeneracy of states with opposite spins and opposite crystal momenta, while the states at a given crystal momentum can be spin split. As a result, the crystal can develop a net spin polarization in a non-equilibrium, current-carrying state. When these spin Hall or Edelstein effects occur at an inversion-asymmetric interface between a paramagnet and a ferromagnet, or inside a ferromagnetic crystal that lacks inversion symmetry, they can induce a spin-orbit torque (SOT) in the ferromagnet \cite{Sinova2015,Hellman2016}. The charge to spin conversion efficiency driving the SOT can outperform that of the STT and is explored as a prospect writing mechanism for future fast MRAMs. A particularly large charge to spin conversion efficiency is expected to occur in time-reversal symmetric topological insulators (TIs) whose 2D surfaces host Dirac cones with the spins locked perpendicular to the 2D momenta \cite{Mellnik2014,Fan2014a,Fan2016}. The ultimate charge to spin conversion efficiency would then occur in 1D surface states of 2D TIs, the so called quantum spin Hall states \cite{Roth2009}, with opposite electron spins locked to the opposite crystal momenta at a given 1D edge.
\begin{figure}[h]
\sidecaption
\includegraphics[width=0.9\linewidth]{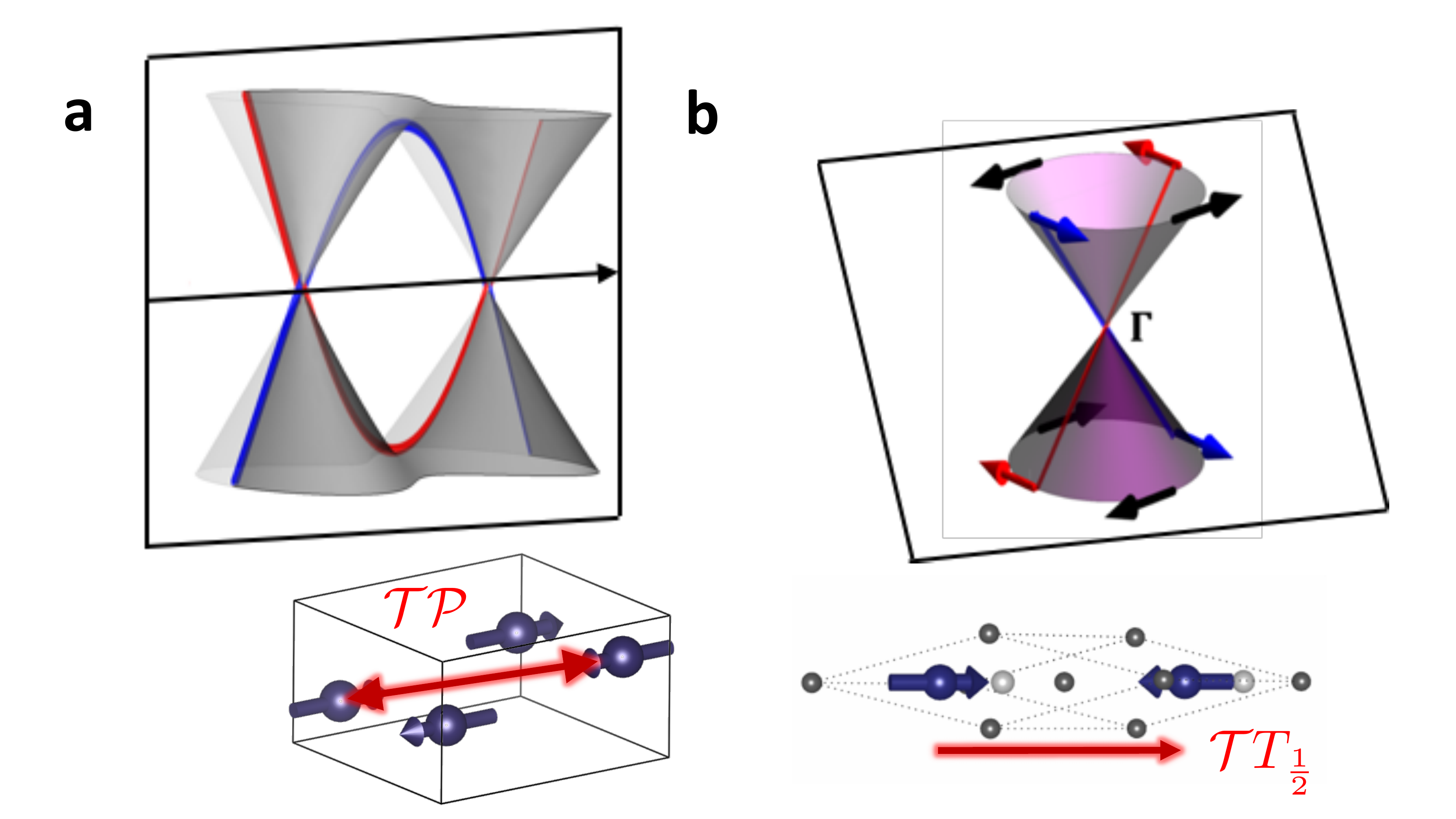}
\caption{\textbf{Concepts of topological antiferromagnetic structures in crystal momentum space.} (a) The antiferromagnetic effective time reversal symmetry $\mathcal{TP}$ combining time reversal and spatial inversion can protect the Dirac semimetal state. The exemplar crystal structure of orthorhombic CuMnAs is shown in the inset. (b) The antiferromagnetic effective time reversal symmetry $\mathcal{T}T_{\frac{1}{2}}$ combining time reversal and half-unit cell translation can protect the topological insulator. The GdPtBi candidate is drawn in the inset. 
}
\label{fig:3}    
\end{figure}

A direct counterpart of the SOT is observed in AFs that break the $\cal{T}$ symmetry and the $\cal{P}$ symmetry but have the combined $\cal{PT}$ symmetry. In these antiferromagnetic crystals, a global electrical current induces a staggered non-equilibrium spin polarization that is commensurate with the staggered equilibrium N\'eel order \cite{Zelezny2014}. The phenomenon was used to demonstrate, in combination with AMR, experimental antiferromagnetic memory devices with electrical writing and readout \cite{Wadley2016}. Because of the THz-scale antiferromagnetic resonance, compared to the GHz-scale ferromagnetic resonance, SOT switching in antiferromagnetic memory devices was demonstrated with writing current pulses as short as 1~ps \cite{Olejnik2017a}. Moreover, since it is again the $\cal{PT}$-symmetry that enables the antiferromagnetic SOT, it is potentially compatible with the large topological AMR  \cite{Smejkal2016}.

Another class of AFs, with no symmetry counterparts in ferromagnets, has the combined ${\cal{T}}T_{\frac{1}{2}}$-symmetry, where $T_{\frac{1}{2}}$ is the translation by a half of the magnetic unit cell. This symmetry allows in principle for TIs with spin-momentum locked surface states, despite the breaking of the $\cal{T}$ symmetry by the magnetic order, as illustrated in Fig. \ref{fig:3}(b) \cite{Mong2010}.  Overall, an antiferromagnetic order can occur in 1421 magnetic space groups out of which only 275 allow also for the ferromagnetic order. Similarly, the 122 magnetic point groups are all compatible with the antiferromagnetic order out of which only 31 also support ferromagnetic states. This not only underlines why antiferromagnetism is more common than ferromagnetism and spans the whole range of materials from insulators to superconductors but also highlights how much is the symmetry and topology playground enlarged by including AFs. We are just beginning to unravel what new spintronics phenomena and functionalities this may offer. In Section 2 we give a brief overview of symmetry and topology concepts in condensed matter systems. In Section 3 we discuss in more detail antiferromagnetic topological semimetals, Chern insulators, and TIs. Topological spintronic phenomena in AFs are reviewed in Section 4 and Section 5 gives a brief summary of the Chapter.

\section{Antiferromagnets: symmetry and topology}
The phases of matter can be classified by Landau symmetry breaking mechanism. In an AF, N\'{e}el vector breaks the rotational symmetry present in the paramagnetic state. The order parameter is the N\'{e}el vector and SOC determines the particular direction(s) of the magnetic moments in crystal, the so-called easy axes, corresponding to the lowest MAE. While certain AFs tends to be to a very good approximation isotropic Heisenberg magnets, e.g. Mn$_{\text{2}}$Au AF can have in-plane MAE at the level of 10\,$\mu$eV per formula unit \cite{Shick2010,Bodnar2017}, noncollinear Mn$_{3}$Sn AF was reported to have MAE of 0.1\,eV per formula unit \cite{Sandratskii1996,Duan2015}, and a giant MAE of 10\,meV per formula unit was predicted in noncollinear IrMn$_{\text{3}}$ \cite{Szunyogh2009}.

The discoveries in the 1980s in the theory of superfluid vortices, QHE, or dislocation defects revealed an additional label of the phases different from the symmetry, based on topology \cite{david1998topological}. The phases can be characterized by an integer topological index which does not change upon continuous transformations of the Hamiltonian and thus supports the relative robustness of a topological phase. For instance, the topological invariants in TIs are the $Z2$ indices, which in simple centrosymmetric non-magnetic TIs are related to counting the number of parities at time-reversal invariant crystal momenta \cite{Hasan2010}. Dirac semimetals (DSMs), such as graphene, are protected by the vorticity around the Dirac point \cite{bernevig2013topological}. The DSMs protected by crystalline symmetry can be assigned a topological index by subtracting this crystalline symmetry eigenvalues of the conduction and valence band along the line in Brillouin zone (BZ) invariant under this symmetry. Red and blue lines in Fig. \ref{fig:3}(a) illustrate the symmetry eigenvalues with an opposite sign. 

\begin{figure}[h]
\sidecaption
\includegraphics[width=1\linewidth]{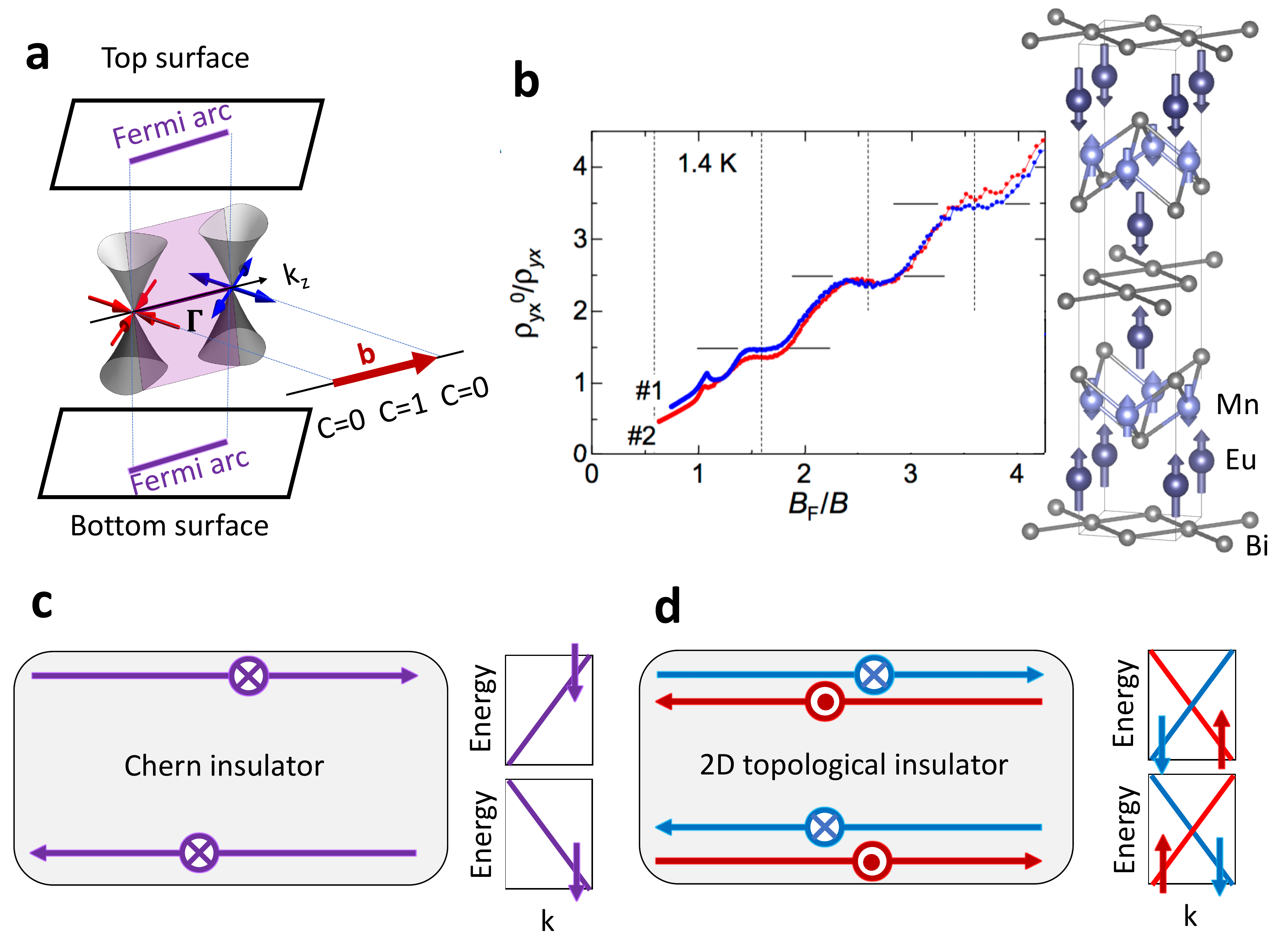}
\caption{\textbf{Topological edge states and transport.} (a) Fermi arcs in Weyl semimetal. (b) The realization of quantum Hall effect as seen in quantized transversal resistivity in EuMnBi$_{\text{2}}$ AF \cite{Masuda2016}. (c) The edge states in Chern insulator. (d) Spin polarised edge states in quantum spin Hall effect. Panel (b) adapted from \cite{Masuda2016}.}
\label{fig:8}       
\end{figure}

We note that TIs and semimetals represent a symmetry protected topological order \cite{Hermanns2017}. For instance, AF DSMs or TIs require the presence of the $\mathcal{PT}$ or $\mathcal{T}T_{1/2}$ symmetry as illustrated in Fig. \ref{fig:3}(a,b). In contrast, Weyl semimetals (WSM) and Chern insulators are protected by Chern numbers. WSM can be realized in system with broken $\mathcal{PT}$ symmetry, while Chern insulators materialize in systems with broken $\mathcal{T}$ symmetry as we explain further. The slices of constant wavevector component $k_{z}$ between two Weyl points in simple model WSM \cite{Vafek2013} (with $\mathcal{PT}$ symmetry broken by breaking $\mathcal{T}$ symmetry, Fig. \ref{fig:8}(a)) can be thought of as Chern insulators, whose examples are QHE or quantum anomalous Hall effect (QAHE). If we use the Bloch ansatz, the Chern number of a $k_{z}$=const. plane reads,
\begin{equation}
\mathcal{C}= \frac{1}{2\pi} \int dk_{x}dk_{y}b_{z}(\textbf{k}),
\end{equation}
where the Berry curvature has a meaning of an emergent magnetic field $\textbf{b}$ in the crystal momentum space and quantifies the underlying topology of the wavefunction \cite{bernevig2013topological}:
\begin{equation}
\textbf{b}(\textbf{k})=-\text{Im} \langle\partial_{\textbf{k}}u(\textbf{k}) \vert \times \vert \partial_{\textbf{k}}u(\textbf{k}) \rangle.
\label{Eq_Berry}
\end{equation}
The topological index of a Weyl point can be defined as a Chern number of a closed surface surrounding the Weyl point, which can be calculated due to the Gauss theorem as a difference between two Chern numbers along the line connecting the Weyl points:
\begin{equation}
\mathcal{Q} = \mathcal{C}(k_{z,W}+\delta)-\mathcal{C}(k_{z,W}+\delta)=\frac{1}{2\pi}\int_{\delta S}d^{2}k \; \textbf{n}\cdot \textbf{b}(\textbf{k}).
\label{WP}
\end{equation}
Here $\delta S$ is a small sphere surrounding the Weyl point at $k_{z,W}$, $\textbf{n}$ is the surface normal vector, and $\mathcal{C}$ is the Chern number of the plane slightly below and above the Weyl point  $k_{z,W}\pm\delta$. Thus the Chern number is nonzero along the $k_{z}$ between the two Weyl points and zero outside as marked in Fig. \ref{fig:8}(a). 

The Berry curvature acts as a source and sink at the Weyl points as we illustrate in Fig. \ref{fig:8}(a) and consequently the two Weyl points along $k_{z}$ have $\cal{Q}$ $=+1$, and $-1$. 
Topological phases in the crystal momentum space are very often accompanied by quantized or almost quantized and low dissipation transport properties and nontrivial surface states \cite{Hasan2010,Wang2017b}. In Fig. \ref{fig:8}(b) we show the quantized Hall plateaus in EuMnBi$_{\text{2}}$ AF \cite{Masuda2016}. The Chern insulator exhibits quantized Hall conductivity:
$
\sigma_{xy}=\frac{e^{2}}{h}\mathcal{C}.
$
On the other hand, quantum spin Hall effect (QSHE) in the 2D TI shows a quantized spin Hall conductivity
$
\sigma_{xy}^{S}=2\frac{e^{2}}{h}.
$
In a Chern insulator, chiral edge states arise (single spin polarized electrons), which are in fact 1D Weyl fermions of a given chirality \cite{Armitage2017}, as we show in Fig. \ref{fig:8}(c).
In 2D TIs, helical edge states are observed (two counter-propagating perfectly polarized currents with opposite spins) as we illustrate in Fig. \ref{fig:8}(d). In contrast, a WSM exhibits surface BZ Fermi arcs for the constant $k_{z}$ in-between Weyl points as we show in Fig. \ref{fig:8}(a). The stacked Fermi arcs from chiral edge states of QAHE subsystems yield almost quantized Hall conductivity:
$
\sigma_{xy}=\frac{e^{2}}{h}\frac{\delta k_{W}}{\pi},
$
where $\delta k_{W}$ is the distance between the Weyl points in the BZ. In Section 3 we will discuss in detail possible realizations of topological semimetals, Chern insulators, and TIs in AFs.

\subsection{Magnetic symmetry and spintronics effects}
\begin{figure}[h]
\sidecaption
\includegraphics[width=1\linewidth]{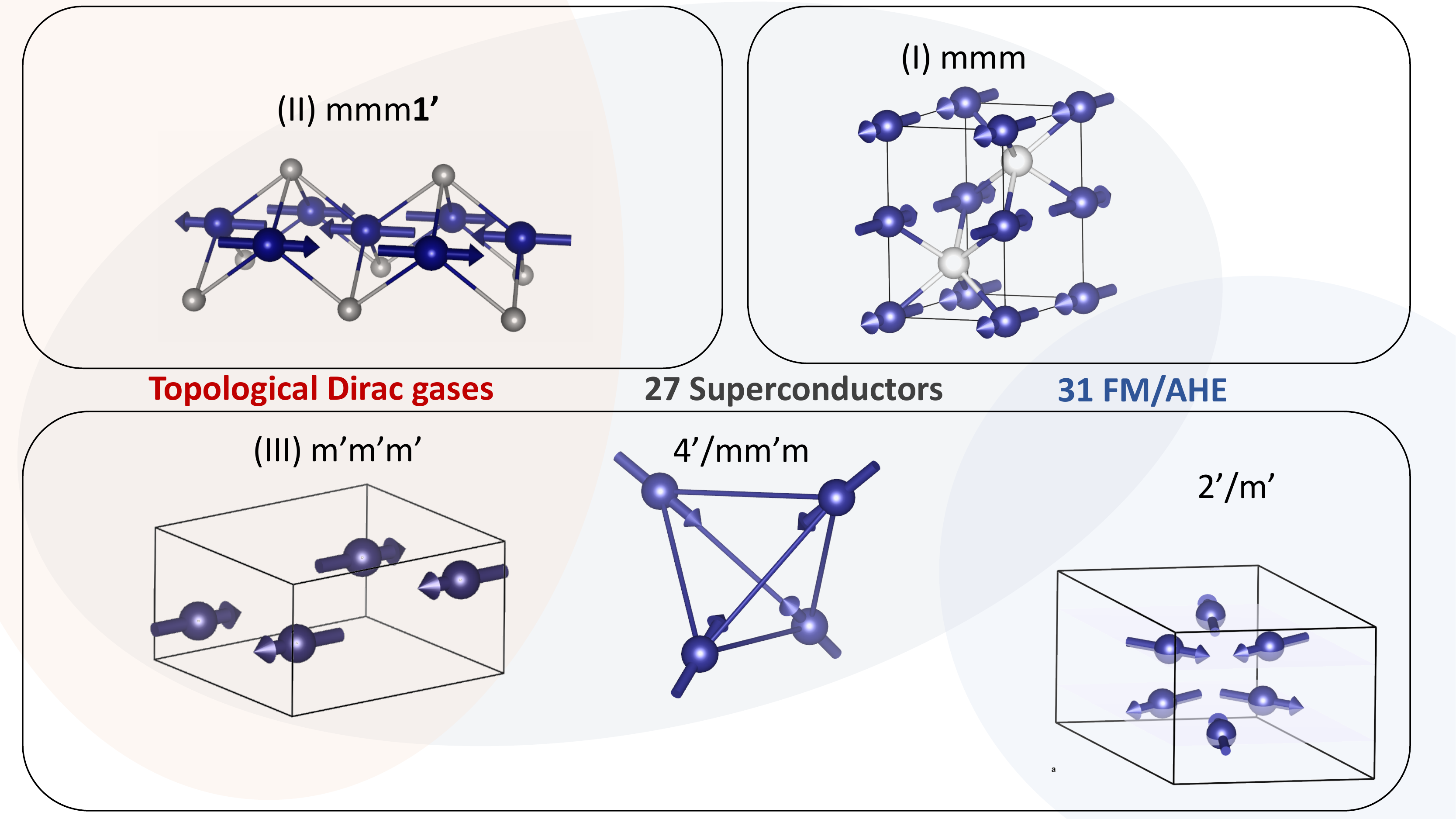}
\caption{\textbf{Classical magnetic point groups (MPGs) and exemplar AFs.} (I) Colorless MPG example of a layered AF MnTe. (II) Grey MPG and zig-zag antiferromagnet FeSe. (III) Black and white MPG. Three different types (from left): $\mathcal{PT}$ AF CuMnAs, centrosymmetric AF with 3Q magnetic order and prohibited anomalous Hall effect (AHE) - IrMn\cite{Kohn2013} or certain pyrochlore AF\cite{Wan2011}, and centrosymmetric AF with nonzero AHE, Mn$_{\text{3}}$Ge. The three colors represent overlap of the antiferromagnetic symmetries allowing for Dirac quasiparticles, superconductivity, and AHE.}
\label{fig:5} 
\end{figure}

The potential presence of long-range ordered antiferromagnetic textures and spintronics effects in AFs is determined by magnetic symmetries.
Magnetic point groups (MPGs) \cite{bradley2010mathematical} are obtained from the ordinary point groups by adding an additional antiunitary operation $\mathcal{T}$ whose application reverses the direction of magnetic moments. Antiferromagnetic order can be in principle found in all 122 MPGs. The colorless MPGs, the so called category I  \cite{bradley2010mathematical} (see an example of a MPG I in Fig. \ref{fig:5}), are those which do not contain the operation $\mathcal{T}$ at all and there are 32 of them which is the same number as the number of nonmagnetic classical point groups. The grey magnetic point groups (category II \cite{bradley2010mathematical}) contains $\mathcal{T}$ as an element of the magnetic symmetry group. There is also 32 of them obtained from the category I by adding the $\mathcal{T}$ operation.  Antiferromagnetic order may appear in this category since the point groups can be obtained from the magnetic space groups by removing all nontrivial unit cell translations. Thus antiferromagnetic sublattices connected by a combination of nontrivial translation $T$ and time reversal fall into this category and we show in Fig. \ref{fig:5}(MPG II) exemplar FeSe AF structure with the $\mathcal{T}T_{\frac{1}{2}}$ operation. Finally, the category III black and white MPGs contain $\mathcal{T}$ only in a combination with another point group symmetry (mirror, or rotation). There are 58 of them and we show in Fig. \ref{fig:5}(MPG III) three antiferromagnetic examples (from left):  A DSM AF crystal\cite{Smejkal2016}, a WSM AF crystal\cite{Wan2011} and a WSM AF crystal with a nonzero AHE \cite{Yang2016c}. 

The form of spintronics linear response tensors is obtained by the application of magnetic symmetries. The Neumann principle states that \textit{any physical observable of a system must exhibit symmetry of the point group of the system} \cite{bradley2010mathematical}. A special role is played by $\mathcal{T}$ and $\mathcal{P}$ symmetries which define the basic transformation properties of tensors as shown in Tab. \ref{tab1}. 
\begin{table}[htbp]
\begin{center}
\begin{tabular}{|l|l|l|l|l|l|l|l|l|}
\hline
Tensor rank & \multicolumn{4}{c|}{Even (scalar, matrix)} &   \multicolumn{4}{c|}{Odd (vector, 3rd rank)}   \\ \hline
Time reversal $\mathcal{T}$ & \multicolumn{2}{c|}{+} &  \multicolumn{2}{c|}{-} &   \multicolumn{2}{c|}{+} & \multicolumn{2}{c|}{-}   \\ \hline
Spatial inversion $\mathcal{P}$ & + (polar) & - (axial) & + (polar) & - (axial) & - (polar) & + (axial) & - (polar) & + (axial) \\ \hline
$\mathcal{PT}$ & + & - & - & + & - & + & + & - \\ \hline \hline
% & AMR &  & AHE &  &  & & $\textbf{j}$, $\textbf{k}$  & LSC  \\ \hline
Exemplar tensor & AMR &  & $\sigma_{ij}^{AHE}$ &  & & $\sigma_{ij}^{S}$ &  & $\textbf{b}$\\ \hline
\end{tabular}
\end{center}
\caption{Spatial inversion and time reversal transformations of tensors.}
\label{tab1}
\end{table}
For the conductivity $\sigma_{ij}$ and spin Hall conductivity $\sigma_{ij}^{S}$ analysis, it is sufficient to use the magnetic Laue group \cite{Seemann2015}. $\sigma_{ij}$ and $\sigma_{ij}^{S}$ do not change sign under spatial inversion and thus this symmetry can be omitted leading to only 32 magnetic Laue groups to investigate \cite{Kleiner1966}. In contrast, the SOT torkance tensor $\tau _{ij}$ changes sign under spatial inversion and thus non-centrosymmetric lattice sites are required and all 122 MPGs have to be considered \cite{Wimmer2016a,Zelezny2017}. This procedure leads to the conclusion that in the MPG from category I, and III there are in total 31 MPGs which allow for uncompensated moments, ferrimagnetism, ferromagnetism and also a nonzero AHE conductivity $\sigma_{ij}^{AHE}$. In Fig. \ref{fig:5}(MPG III) - right panel we show a corresponding example of the non-collinear AF structure of Mn$_{\text{3}}$Ge. 
\begin{figure}[h]
\sidecaption
\includegraphics[width=\linewidth]{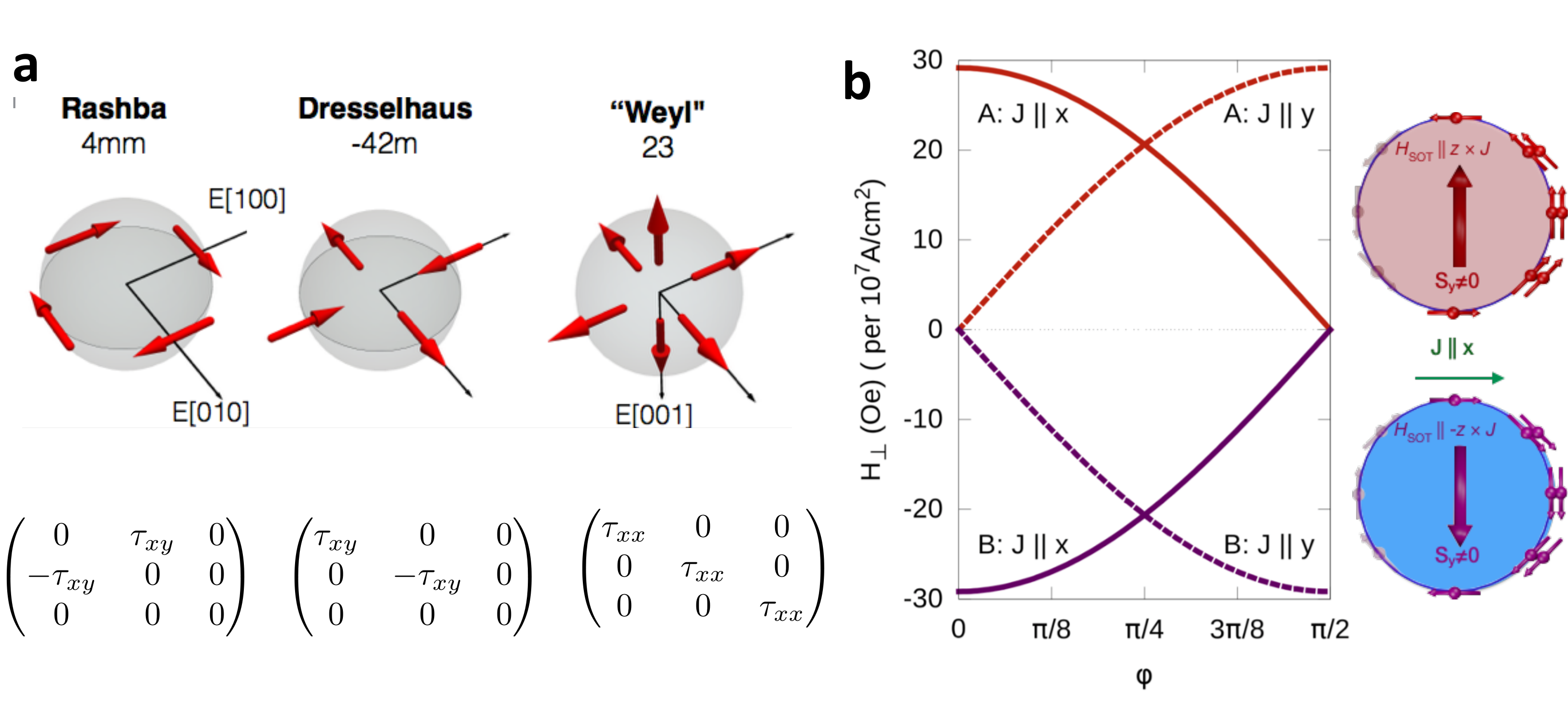}
\caption{\textbf{The symmetry of spin-orbit fields.} 
(a) All possible nonequilibrium spin polarization at noncentrosymmetric positions in crystals can be decomposed into combinations of Rashba, Dresselhaus, or Weyl spin texture. (b) Example of realistic calculation of the spin-orbit fields induced by electric current in tetragonal CuMnAs. The symmetry of the fields was confirmed in experiment \cite{Wadley2016}.  Panel (b) adapted from \cite{Wadley2016}.}
\label{fig:6}       
\end{figure}

The torkance tensor is defined by $\textbf{t}=\tau \textbf{E}$ \cite{Freimuth2014a}, where $\textbf{t}$ is a torque generated by an applied electric field $\textbf{E}$. The Onsager reciprocal for SOT is the inverse SOT obtained by interchanging the perturbation and response and both effects can be desribed by the torkance tensor \cite{Freimuth2015}.  SOT can be decomposed into an even part $\textbf{t}_{FL}$ and an odd part $\textbf{t}_{AL}$:  $\textbf{t}=\textbf{t}_{FL}+\textbf{t}_{AL}$. There are 21 non-centrosymmetric point groups which allow for a global or local current-induced nonequilibrium spin polarisation, and thus for SOT, in the lowest order that is independent of magnetization \cite{Ciccarelli2016}. Spin-orbit fields generating the SOT can be decomposed into a combination of Rashba, Dresselhaus, and Weyl symmetry as shown in Fig. \ref{fig:6}(a). The AF variants of SOT can be found in AFs with non-centrosymmetric magnetic sublattices connected by crystalline symmetries as we illustrate on the example of the CuMnAs AF in Fig. \ref{fig:6}(b) \cite{Zelezny2014,Wadley2016,Smejkal2016,Bodnar2017}. Here the dominating SOT is driven by a staggered effective current-induced field of Rashba symmetry and the magnetic sublattices are connected via $\mathcal{PT}$ symmetry \cite{Smejkal2016}. Wimmer et al. \cite{Wimmer2016a} list forms of torkance tensors for all 122 MPGs.  

There are also 21 MPGs which contain combined $\mathcal{PT}$ symmetry and can host Dirac quasiparticles as we explain in Section 3. 
Other effects can be treated analogically. For instance, the magneto-optical Kerr effect \cite{Feng2015} and anomalous Nernst effect \cite{Ikhlas2017} were predicted and observed in non-collinear AFs as well. Also, this scheme can be applied to the layer-resolved quantities in heterostructures, e.g. the layer-resolved conductivity \cite{Seemann2015}. 

Finally, AF order can coexist with superconductivity as we illustrate in Fig. \ref{fig:5} by the grey shaded ellipse and examples are the iron-based superconductors \cite{Wang2016e}. More complicated magnetic structures such as spin spirals, spin density waves, and skyrmions may not allow to be described completely in the classical MPG framework. Gopalan and Litvin \cite{Gopalan2011} have shown the possibility of new hidden symmetries, an example being local roto-inversion. This operation does not rotate the whole crystal but just a finite subset while unchanging its MPG. The novel magnetic counterpart symmetries might be also relevant for the complete description of antiferromagnetic structures.

\subsection{Electronic structure and band touchings}
Antiferromagnetic exchange interactions arise as a result of the complex interplay among the electrons. Different types of exchange interactions are possible, e.g., direct, indirect, superexchange, or itinerant exchange. The electronic structure of AFs is thus very often complicated and requires the inclusion of correlation and many particle effects. A realistic insight into the electronic structure and the existence of the antiferromagnetic phase can be determined by the density functional theory (DFT). In DFT, the interacting many-particle problem is mapped onto non-interacting electrons in an effective Kohn-Sham potential \cite{Kohn1965}. Hohenberg and Kohn have shown that the ground-state properties of the effective electronic gas are uniquely determined by the electronic density \cite{Hohenberg1964}. The reformulation of the problem as a variational one tremendously decreases the macroscopic $\sim$10$
^{23}$ degrees of freedom to just 3 - the spatial coordinates of the electronic density. In a magnetic system with a strong relativistic SOC, the magnetic relativistic "spin-only" (neglected diamagnetic effects \cite{strange1998relativistic}) Kohn-Sham-Dirac Hamiltonian reads \cite{strange1998relativistic}:
\begin{equation}
H_{\text{KSD}}=c\alpha \cdot \textbf{p}+\beta mc^{2}+V^{\text{eff}}(\textbf{r})-\textbf{m}(\textbf{r})\cdot\textbf{B}^{\text{eff}}(\textbf{r}),
\label{KSD}
\end{equation}
where $\textbf{p}=-i\hbar \nabla$ is a momentum, $V^{\text{eff}}$ and $\textbf{B}^{\text{eff}}$ is the spin independent part of the potential and the exchange-correlation magnetic field, $\alpha, \beta$ are 4x4 Dirac matrices, $\textbf{m}$ is spin density, and the electronic density is obtained from $n(\textbf{r})=\sum \Psi_{i}^{\dagger}(\textbf{r})\Psi_{i}(\textbf{r})$ \cite{strange1998relativistic}. 
The Kohn-Sham potential is not known exactly and has to be approximated, e.g., by the local density approximation (LDA) or generalized gradient approximation (GGA) \cite{prasad2013electronic}. 
The set of equations for the electronic wavefunctions and Kohn-Sham potential is solved iteratively. The procedure yields ground state wavefunctions and Hamiltonian from which other quantities, e.g., linear response coefficients can be calculated. For instance in Fig. \ref{fig:9}(b) we show electronic bands of the orthorhombic AF CuMnAs as calculated within GGA \cite{Smejkal2016}.

AF systems are often correlated and disordered. Electronic correlations can be treated within DFT+Hubbard U, or DFT+dynamical mean field theory (DMFT). We show the generalized band structure of an AF BaFe$_{\text{2}}$As$_{\text{2}}$ calculated by DFT+DMFT in Fig. \ref{fig:FeSe}(b). The effects of disorder can be captured by the supercell technique \cite{prasad2013electronic} or coherent potential approximation \cite{prasad2013electronic} as was demonstrated, for instance, for disordered Mn$_{\text{2}}$Au AF \cite{Bodnar2017}.

Symmetries impose constraints on the electronic spectrum, including the existence and protection of band touchings. The palette of quasiparticles in solids is more rich than the three types of high energy physics excitations: Weyl, Dirac and Majorana fermions \cite{Vafek2013}. This is because of the more complex crystalline symmetries that are not present in the high energy vacuum \cite{Wieder2016,Bradlyn2016}. In the next Section we illustrate how the effective Hamiltonian arises for Dirac and Weyl quasiparticles in an AF. 

\section{Topological antiferromagnetic phases}
Topological magnetic phases can be found in heterostructures with antiferromagnetic elements as well as in bulk AFs. In Tab. \ref{tab2}, we list promising topological AFs for spintronics together with the status of theoretical predictions and experimental observations of topological state or spintronics effects. 

\begin{table}
\caption{Topological antiferromagnets}
\label{tab:1}      
\begin{tabular}{p{1.5cm}p{3.5cm}p{1.5cm}p{1.5cm}p{5.5cm}}
\hline\noalign{\smallskip}
AF & Phase & T$_{N}$ & Space group  & Representative effect \\ \hline
\noalign{\smallskip}\svhline\noalign{\smallskip}

FeSe & QSHE \cite{Wang2016e} &  & P4/nmm & Superconductivity \\ \hline
GdPtBi & \textit{TI}\cite{Mong2010,Li2011,Li2015}/Weyl \cite{Hirschberger2016} & 9 \cite{Suzuki2016} & F$\overline{4}$3m & Large thermopower \cite{Hirschberger2016}, AHE \cite{Suzuki2016}  \\ \hline
\noalign{\smallskip}\svhline\noalign{\smallskip}
SrMnBi$_{2}$ & Dirac metal\cite{Park2011a} & 290 \cite{Park2011a} & I4/mmm & Angular dependent magnetoresistance \cite{Wang2011e} \\ \hline
CaMnBi$_{2}$ & Dirac metal & 300 \cite{Guo2014} & P4/nmm &  Dirac fermions coupled to magnetism \cite{Guo2014} \\ \hline
EuMnBi$_{2}$ & Dirac metal & 22* \cite{Masuda2016} & I4/mmm & QHE controlled by magnetism \cite{Masuda2016} \\ \hline 
BaFe$_{2}$As$_{2}$ & 2D Dirac metal\cite{Chen2017}  &   143 \cite{Huang2008} & I4/mmm & Superconductivity \\ \hline 
CuMnAs & \textit{Dirac semimetal}\cite{Maca2012,Tang2016,Smejkal2016} & $\sim$400 \cite{Maca2012}  & Pnma &  \textit{TopoMIT, TopoAMR} \cite{Smejkal2016} \\ \hline \hline
X$_{2}$ Ir$_{2}$ O$_{7}$  & Weyl semimetal\cite{Wan2011,Sushkov2015} &  &  Fd$\overline{3}$m & TopoMIT \cite{Tian2015}, \textit{wealth of topo. phases} \cite{Wan2011,Kondo2015} \\ \hline
Mn$_{3}$Sn  & Weyl (semi)metal \cite{Yang2016c,Kuroda2017} & 430 \cite{Nakatsuji2015} & P63/mmc & AHE controlled by magnetic field\cite{Nakatsuji2015} \\ \hline
\end{tabular}
$^a$ Italic font marks theoretical prediction. Normal font marks existing experimental signatures. 
\label{tab2}
\end{table} 

\subsection{Low dimensional Dirac antiferromagnets and superconductors}
Introducing magnetism into TIs is known to modify the spin texture of the Dirac quasiparticles \cite{Xu2012}. Magnetism can couple to TIs either by creating the magnetically doped TI (MTI) \cite{Xu2012}, or by proximity coupling between the TI and magnetic order in heterostructures \cite{Katmis2016}. AF order was shown to increase the critical temperature of the adjacent MTIs in the MTI/AF CrSb/MTI heterostructure \cite{He2016}. Improved performance of the SOT in terms of larger spin Hall angles and lower critical currents was achieved in the TI/ferrimagnetic CoTb alloys \cite{Han2017} with the AF coupled \cite{Finley2016} Co and Tb sublattices. 

\begin{figure}[h]
\sidecaption
\includegraphics[width=1\linewidth]{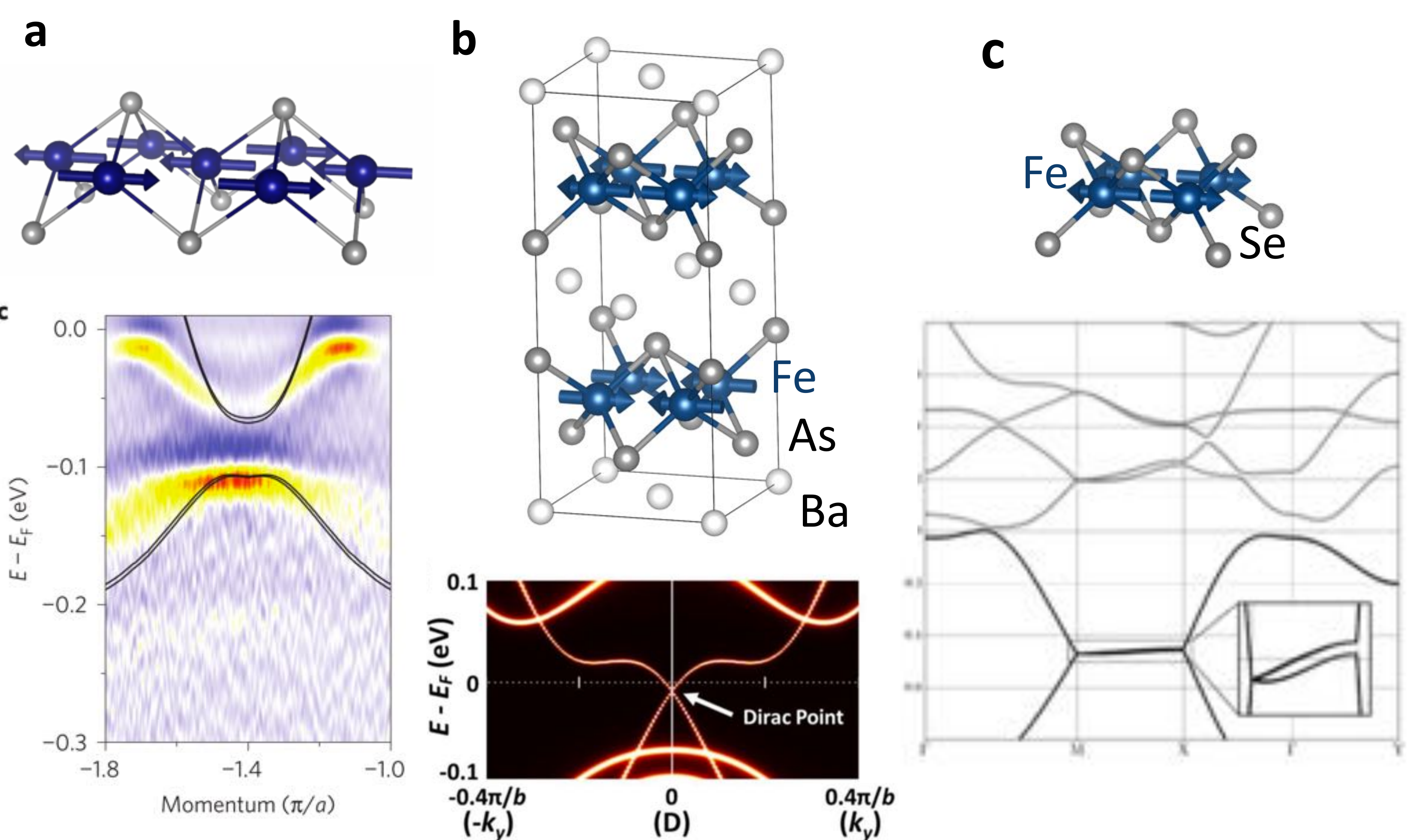}
\caption{\textbf{Dirac quasiparticles in iron-based AF superconductors.} (a) Realization of QSHE in a monolayer of FeSe AF. Angle resolved photoemission spectroscopy data overlayed with \textit{ab initio} bands \cite{Wang2016e}. 
(b) Topological quasi-2D Dirac quasiparticles in BaFe$_{\text{2}}$As$_{\text{2}}$ AF. (c) Single Dirac cone at M point in FeSe monolayer AF with a stripe order.  Panel (a) adapted from \cite{Wang2016e}, panel (b) from \cite{Chen2017}, and panel (c) from \cite{Young2016}.}
\label{fig:FeSe}       
\end{figure}

The first TI AF was predicted in systems with combined $\mathcal{T}T_{1/2}$ symmetry \cite{Mong2010}. GdPtBi AF was suggested as a possible candidate, as we show in Fig. \ref{fig:3}(b). GdPtBi was up to date not confirmed as a TI presumably due to the low resolution of data obtained by the angle resolved photoemission spectroscopy (ARPES) \cite{Liu2011f}. However, signatures of the coexistence of a 2D TI (Fig. \ref{fig:8}(c)), and a superconducting state in hole-doped and electron-doped antiferromagnetic monolayers of
 FeSe were demonstrated \cite{Wang2016e}. FeSe belongs to the metallic building block of the iron-based high-T$_{\text{C}}$ superconductors. Remarkably, the combined effects of SOC, substrate strain, and electronic correlations can induce band inversion and QSHE edge states, as we show in Fig. \ref{fig:FeSe}(a) \cite{Wang2016e}. Creating a p-n junction across FeSe and attaching two ferromagnetic electrodes can generate Majorana zero modes at the interfaces \cite{Tsai2016}. Majorana states are considered for a possible use in quantum computing \cite{Beenakker2016}.

One of the first systems explored for observing Dirac quasiparticles in condensed matter beyond graphene were the SrMnBi$_{\text{2}}$ type AFs. The electronic structure of these systems is governed by the quasi-2D square Bi planes. The Bi states create close to the Fermi level massive Dirac quasiparticles. High mobilities, and Fermi velocities, and pseudospin structure of wavefunctions are reminiscent of graphene properties. In contrast to graphene, however, the quasiparticles are highly anisotropic with anisotropy factor of $\sim$ 8 \cite{Wang2011e}. Several of these types of AFs were reported in recent years including SrMnBi$_{\text{2}}$ \cite{Wang2011e}, or CaMnBi$_{\text{2}}$ \cite{Guo2014}. The systems belong to the 112-type pnictides where the antiferromagnetism and Dirac quasiparticles might coexist also with superconductivity. Related systems, e.g. YbMnBi$_{\text{2}}$, were inconclusively \cite{Armitage2017} reported to be either WSM \cite{Borisenko2015,Chinotti2016} or DSM \cite{Wang2016k,Chaudhuri2017}. Most likely the collinear AF order cants in, e.g., the (Sr,Yb)MnBi$_{\text{2}}$ alloy \cite{Liu2017} where the double band degeneracy breaks and Weyl points might emerge. Despite many recent studies, more accurate and detailed measurements are needed to reveal the detailed nature of Dirac quasiparticles in these systems. Finally, in the sister compound, EuMnBi$_{\text{2}}$ AF, the half-integer QHE was reported controllable by the strength of an external magnetic field, as we showed in Fig. \ref{fig:8}(b). EuMnBi$_{\text{2}}$ contains at very low temperatures two antiferromagnetic sublattices. The presence of QHE was linked to the confinement of the massive Dirac quasiparticles by the spin-flop at the Eu AF sites \cite{Masuda2016}.

Recently, the high temperature superconducting 122-type pnictide XFe$_{\text{2}}$As$_{\text{2}}$ (X=Ba, or Sr) AFs showed signatures of topological Dirac quasiparticles in the infrared spectra in high magnetic fields. In Fig. \ref{fig:FeSe}(b) we show the state-of-the-art DFT+DMFT calculation of quasi-2D Dirac cones close to the Fermi level which are consistent with the observed Landau level spectra and density of states \cite{Chen2017}. 
When the Fermi states are dominated by Dirac quasiparticles, the topological semimetal is achieved. Electron filling enforced semimetals with a single Dirac cone were predicted theoretically in 2D model AFs \cite{Young2016}. In Fig. \ref{fig:FeSe}(c) we show the single Dirac cone at the $M$ point in the BZ of the monolayer of the FeSe AF with a stripe order. The quasi-low dimensional systems and heterostructures, however, suffer from fragile magnetism and low critical temperatures. In two following subsection, we describe possible room temperature 3D Dirac and Weyl semimetal AFs.

\subsection{3D Dirac semimetal antiferromagnets}
\begin{figure}[h]
\sidecaption
\includegraphics[width=1\linewidth]{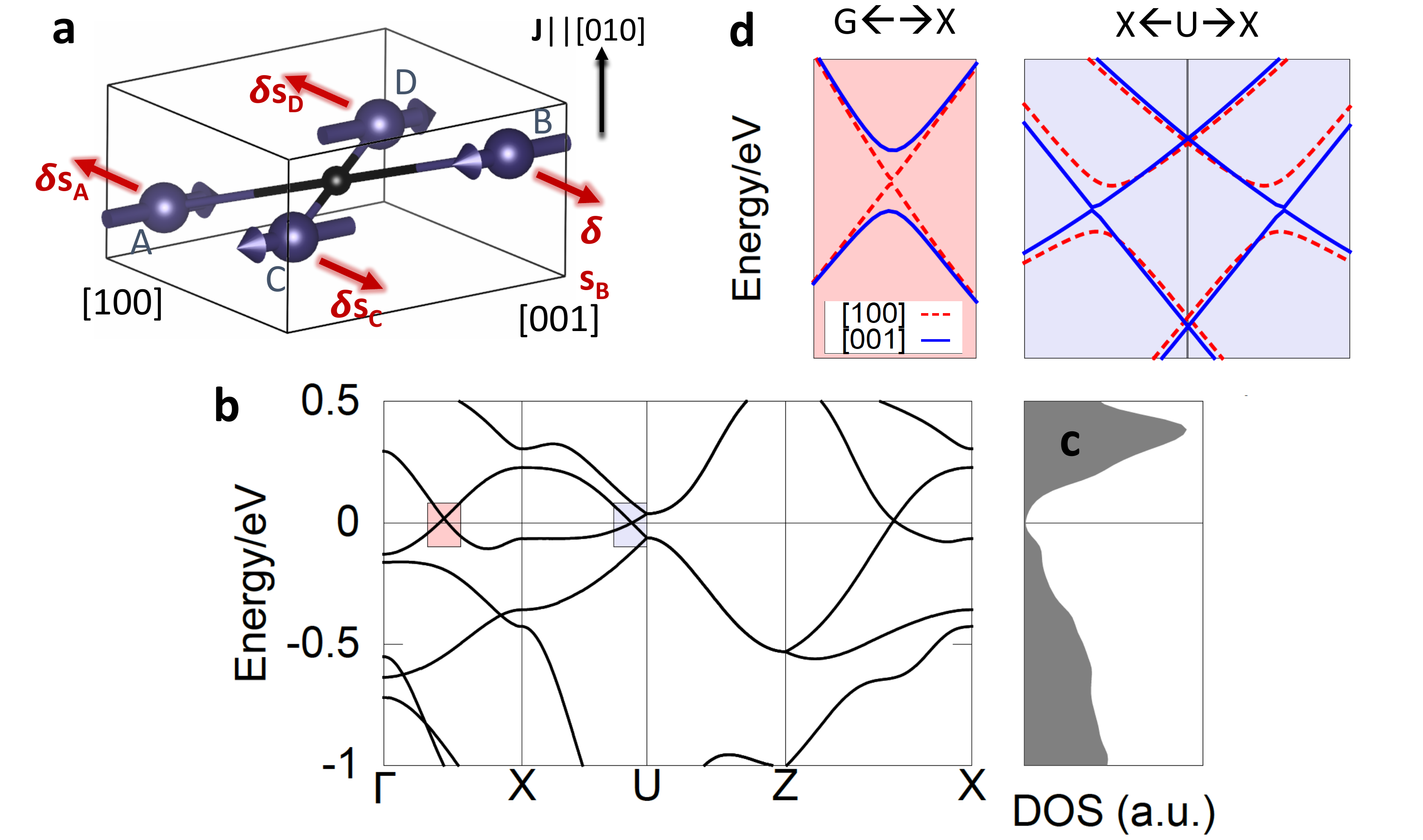}
\caption{\textbf{Antiferromagnetic Dirac semimetals, spin-orbit torques, and metal-insulator transition.} (a) Unit cell of orthorhombic CuMnAs AF with marked $\mathcal{PT}$ symmetry center (black sphere) and nonequilibrium spin polarizations $\delta\textbf{s}$ for current applied along the [010]. (b) Band structure of the CuMnAs AF calculated \textit{ab initio} without SOC. (c) The density of states. (d) Detail of Dirac quasiparticles in CuMnAs as calculated \textit{ab initio} with SOC switched on. Panel (a) adapted from Ref.\cite{Smejkal2017a}, and panels (b-d) adapted from Ref.\cite{Smejkal2016}.}
\label{fig:9}     
\end{figure}
Dirac quasiparticles are allowed in doubly-degenerate bands \cite{bernevig2013topological} realized in systems invariant under $\mathcal{PT}$ symmetry.  
\begin{svgraybox}
The low energy Hamiltonian might maintain an effective Dirac form \cite{Vafek2013,Yang2014a,Smejkal2017}, corresponding to the Dirac Hamiltonian \eqref{KSD} \cite{Burkov2016,Smejkal2017,Armitage2017}: 
\begin{align}
\mathcal{H}_{D}(\textbf{k}) = 
\left(\begin{matrix} 
 \hbar v_{F}\textbf{k}\cdot \boldsymbol\sigma & m \\
m &  -\hbar v_{F}\textbf{k}\cdot \boldsymbol\sigma
\end{matrix}\right).
\label{MDirac}
 \end{align} 
Here $v_{F}$ is the Fermi velocity, $\textbf{k}=\textbf{q}-\textbf{q}_{0}$ is the crystal momentum measured from the Dirac point at $\textbf{q}_{0}$, $m$ is the mass (in units of energy), and $\boldsymbol\sigma$  is the vector of Pauli matrices. The energy dispersion gives massive Dirac cones,
\begin{equation}
E(\textbf{k})=\pm \hbar v_{\text{F}}\sqrt{k_{x}^{2}+k_{y}^{2}+k_{z}^{2}+\left(\frac{m}{\hbar v_{F}}\right)^2}.
\end{equation}
\end{svgraybox}
The mass term can be removed by the presence of an additional  crystalline symmetry,
and in this case $\mathcal{H}_{D}(\textbf{k})$ describes the  
four-fold degenerate band touching \cite{Burkov2016,Smejkal2017} of a 
3D DSM as we show in Fig. \ref{fig:9}(b) on the band dispersion of an antiferromagnetic DSM orthorhombic CuMnAs with a high N\'{e}el temperature of $\sim$ 400\,K \cite{Maca2012}. 

The 3D DSM state cannot occur in ferromagnets because $\mathcal{T}$-symmetry breaking prevents double band degeneracy. However, a topological crystalline 3D DSM was predicted in an AF, namely in the orthorhombic phase of CuMnAs \cite{Tang2016,Smejkal2016}. The unit cell of the orthorhombic CuMnAs contains four Mn sublattices that are connected in pairs by the $\mathcal{PT}$ symmetry \cite{Tang2016,Smejkal2016} as we show in Fig. \ref{fig:9}(a). Although individually the $\mathcal{P}$ and $\mathcal{T}$ symmetries are broken, the preserved combined $\mathcal{PT}$ symmetry ensures the double band degeneracy over the whole BZ. In the calculation with a switched-off SOC (see Fig. \ref{fig:9}(b)), we observe three Dirac points at the Fermi level along the $\Gamma−X$, $X−U$, and $Z−X$ lines which are part of the nodal line protected by the $\mathcal{PT}$ symmetry. The Dirac quasiparticles are 3D as can be seen from the quadratically vanishing DOS at the Fermi level/Dirac point as we show in Fig. \ref{fig:9}(c). In the presence of SOC and for the N\'{e}el order along the [001] axis, the nodal lines become gapped except for the two Dirac points along the $UXU$ line, as we show in Fig. \ref{fig:9}(d). The Dirac points are protected by the non-symmorphic screw axis symmetry $\mathcal{S}ß_{z}=\left\lbrace 2_{z} \vert (\frac{1}{2},0,\frac{1}{2}) \right\rbrace$ \cite{Tang2016,Smejkal2016} and are connected via nontrivial surface states \cite{Tang2016}. The topological invariants and surface states can be linked to the crystalline symmetry protecting the degeneracy and even in the non-magnetic DSM, the surface states are in general less stable than in WSM and strongly depend on the crystalline orientation at the surface termination \cite{Yang2014a,Kargarian2016}. The easy axis in orthorhombic CuMnAs tends to be along [100] according to \textit{ab initio} calculation \cite{Smejkal2016}, however, we will discuss in the next section the possibility of reorienting the N\'{e}el vector.

The orthorhombic CuMnAs AF is an attractive {\it hydrogen atom} for magnetic DSMs induced by band inversion since only a single pair of Dirac points appears near the Fermi level of the {\em ab initio} band structure. However, presumably, the correlation and disorder effects prevented the observation of Dirac quasiparticles in non-stoichiometric CuMnAs to date \cite{Emmanouilidou2017,Zhang2017e}.

\subsection{Weyl semimetal antiferromagnets}
In solids, quite often at least one of the $\mathcal{P}$ or $\mathcal{T}$ symmetries (and also $\mathcal{PT}$) is broken and thus the double band degeneracy is lifted. When the two non-degenerate bands are touching close to the Fermi level a 3D WSM can be formed. 
\begin{svgraybox}
WSM is described by the generalized two-band Weyl Hamiltonian \cite{Burkov2016,Smejkal2017}: 
\begin{equation}
\mathcal{H}_{W}(\textbf{k})=\epsilon_{0}\pm\hbar v_{\text{F}}\left( \textbf{q} - \textbf{q}_{0}  \right)\cdot \boldsymbol{\sigma},
\label{Weyl}
\end{equation}
where the first term corresponds to the tilting of the Weyl cone and $\textbf{k}=\textbf{q} - \textbf{q}_{0} $. Weyl points always come in pairs with opposite topological charges. Dimensionality is important here. Because the Weyl points are 3D objects in the BZ, the effective Hamiltonian uses all three Pauli matrices. Thus any small perturbation expressed without loss of generality as a linear combination of these three Pauli matrices just shifts but not gaps the Weyl point. We illustrate this on the dispersion around the Weyl points for a perturbation of a form $m\sigma_{z}$,
\begin{equation}
E(\textbf{k})=\epsilon_{0}\pm \hbar v_{\text{F}}\sqrt{k_{x}^{2}+k_{y}^{2}+\left(k_{z}+\frac{m}{\hbar v_{\text{F}}}\right)^{2}}.
\label{WSMdisp}
\end{equation}
\end{svgraybox}

WSM states can be found in nonmagnetic, ferromagnetic \cite{Xu2011,Wang2016c} or antiferromagnetic solids, where the $\mathcal{PT}$ symmetry is broken. A WSM state was observed in non-centrosymmetric non-magnetic mono-pnictides of the TaAs type \cite{Xu2015b,Lv2015,Yang2015d}. TaAs is well described by the DFT single quasiparticle picture. Despite numerous predictions, the true magnetic WSM remained for a long time experimentally elusive \cite{Kuroda2017}. The reason is that the magnetic system is very often also strongly correlated, disordered, and the symmetry breaking is provided by the complex collective phenomenon - magnetism. Antiferromagnetic candidates include pyrochlore irridates \cite{Wan2011} like the Eu$_{2}$Ir$_{2}$O$_{7}$ \cite{Sushkov2015}, or YbMnBi$_{2}$ AF which were suggested to be either the WSM \cite{Borisenko2015,Chinotti2016} or the DSM \cite{Wang2016k,Chaudhuri2017}.  

\begin{figure}[h]
\sidecaption
\includegraphics[width=1\linewidth]{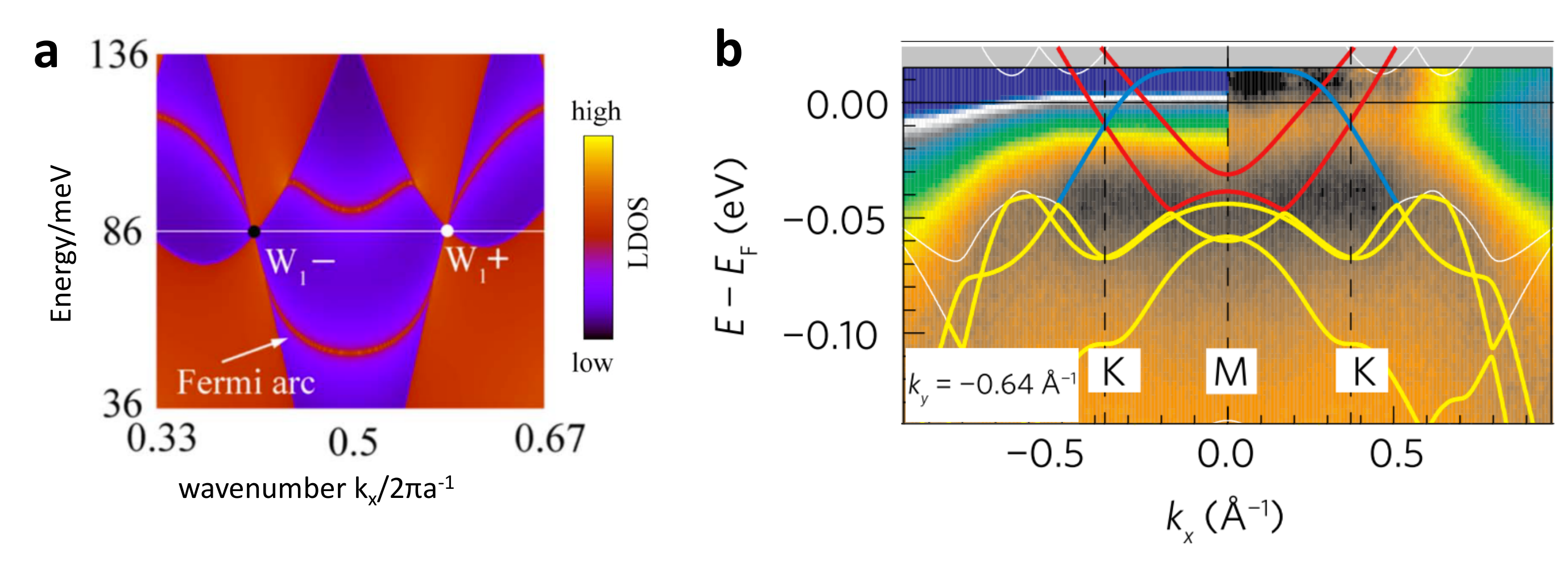}
\caption{\textbf{Antiferromagnetic Weyl semimetal and surface Fermi arcs.} 
(a) \textit{Ab initio} calculation of Fermi arcs in Mn$_{3}$Sn AF. (b) ARPES of Weyl points in Mn$_{3}$Sn close to the Fermi level overlayed with \textit{ab initio} band structure. Panel (a) adapted from Ref. \cite{Yang2016c} and panel (b) from \cite{Kuroda2017}.}
\label{fig:10}       
\end{figure}
Weyl fermions were proposed also in non-collinear AFs Mn$_{3}$Ge and Mn$_{3}$Sn \cite{Kubler2014,Yang2016c} (see Fig. \ref{fig:2}(a)) \cite{Yang2016c}. 
These AFs are potentially appealing for spintronics due to the measured large AHE, established magnetic structure, and N\'{e}el temperatures reaching 365-420 K.
The structure of Mn$_{3}$Ge and Mn$_{3}$Sn crystals is built from stacked kagome planes along the [001] axis, as we show in the right panel in Fig. \ref{fig:5}(MPG III). 
These AFs have a relatively weak magnetic anisotropy, reaching 0.1meV per formula unit for Mn$_{3}$Sn \cite{Sandratskii1996,Duan2015} due to the vanishing second and fourth order MAE of the inverted triangular AF structure on the kagome lattice \cite{Nagamiya1982,Tomiyoshi1982}. The magnetic order can be thus reoriented by low external magnetic fields.  Ref. \cite{Kuroda2017} reports reorientation fields of $\sim$ 200 Gauss. The net magnetic moment reaches 0.005 $\mu$B per unit cell \cite{Tomiyoshi1982}. In spite of the weak anisotropy of the inverted chiral structure, the materials show relatively high stability against thermal fluctuations. Also, a possibility to influence the in-plane chiral AF magnetic structure by a spin-filtering effect was reported \cite{Fujita2017}.

Mn$_{3}$Ge and Mn$_{3}$Sn were predicted to exhibit several different types of Weyl points in their metallic bandstructure coexisting with trivial bands close to the Fermi level \cite{Yang2016c}. The Weyl points found by tracking the Berry curvature in the BZ are tilted, thus of the so-called type-II \cite{Soluyanov2015}. The Fermi arc surface states - the hallmark of a WSM - were predicted by first-principles calculations of the local density of states (LDOS) as we show in Fig. \ref{fig:10}(a) \cite{Yang2016c}. The tilting of the Weyl points does not influence the Berry curvature, however, the electron and hole pockets due to the tilting influence the transport effects, particularly they can renormalize the almost perfectly quantized AHE \cite{Zyuzin2016a}. 
Signatures of these band crossings were reported recently in an ARPES study of Mn$_{3}$Sn AF \cite{Kuroda2017}, as we show in Fig. \ref{fig:10}(b). Weyl band crossings were found along $MK$ or $M'K$ lines presumably depending on the orientation of the triangular magnetic texture stabilized by an external magnetic field \cite{Kuroda2017}. A comparison between DFT and ARPES points towards strong electron correlations in the Mn 3d bands as seen in the strong renormalization of the bands in ARPES in Fig. \ref{fig:10}(b). In the next Section, we describe in detail the AHE in Mn$_{3}$Ge and Mn$_{3}$Sn and its relation to Weyl points. 

\section{Topological antiferromagnetic spintronics effects}
Topological variants of common magnetotransport effects and their novel cousins can potentially offer large signal to noise ratios important for reading signals in spintronics nanodevices.  Again, we will demonstrate that very often the unique antiferromagnetic symmetries are of vital importance for certain topological spintronics effects to occur, such as topological AMR or AHE. 

\subsection{Large magnetoresistance and chiral anomaly}
\begin{figure}[h]
\sidecaption
\includegraphics[width=1\linewidth]{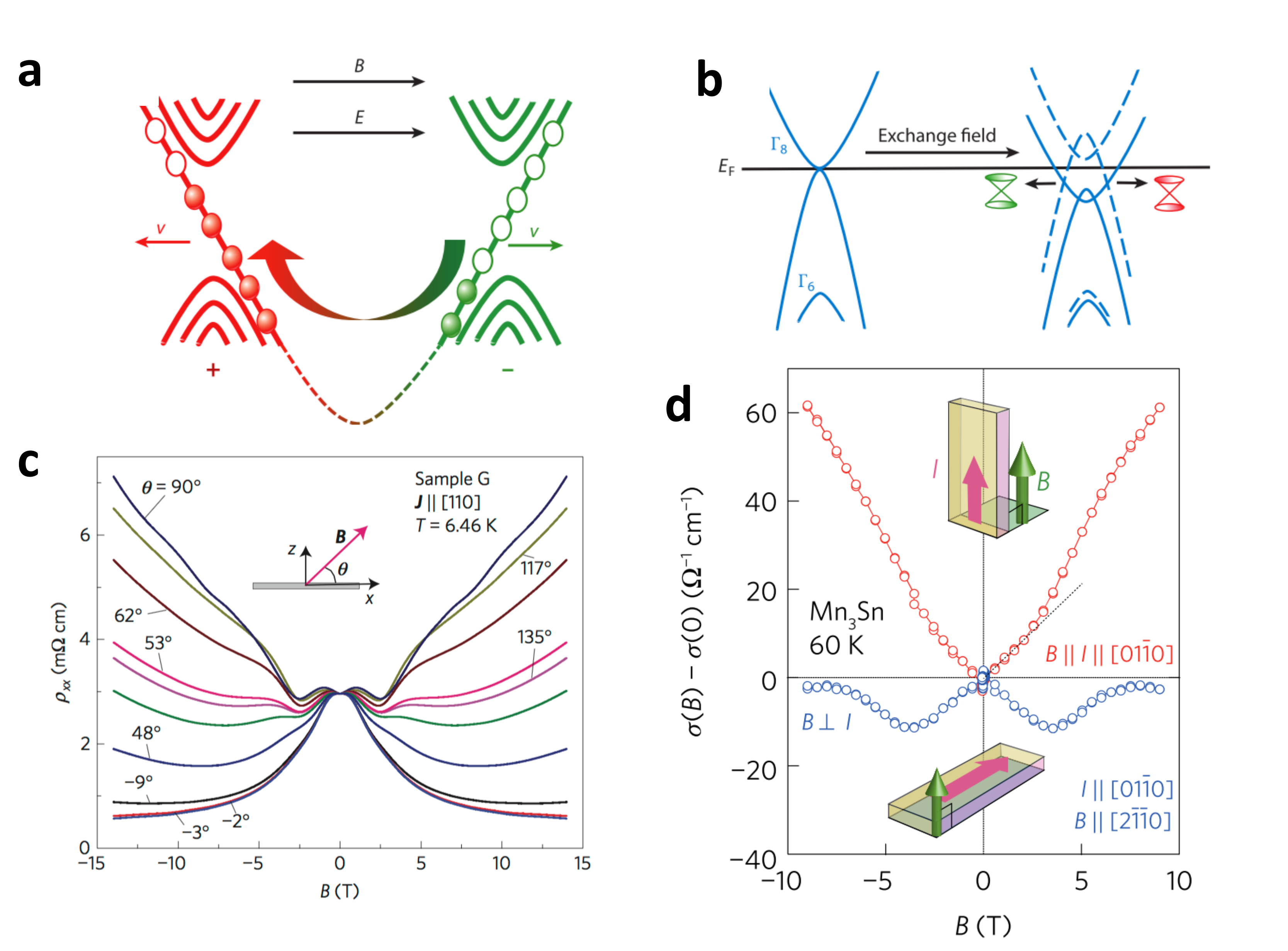}
\caption{\textbf{Possible observation of a chiral anomaly in magnetically induced Weyl semimetal and antiferromagnetic Weyl semimetal.} (a) Cartoon of the chiral anomaly principle. (b) A magnetically induced Weyl semimetal in GdPtBi. (c) Observed negative magnetoresistance in GdPtBi. (d) Observed positive magnetoconductance in the Mn$_{3}$Sn AF. Panels (a,b) adapted from \cite{Yan2016a}, panel (c) from \cite{Hirschberger2016}, and panel (d) from \cite{Kuroda2017}.}
\label{fig:11}       
\end{figure}
Application of a magnetic field perpendicular to the current flow results in a positive magnetoresistance, defined as 
\begin{equation}
\text{MR}=\frac{\rho(B)-\rho(0)}{\rho(0)},
\end{equation}
where $\rho(B)$ is the resistivity in a magnetic field $B$, as commonly observed in metals, semimetals, and semiconductors. However, in a topological semimetal with Weyl quasiparticles, a negative magnetoresistance might occur, attributed to a chiral anomaly. In the original proposal for a consensed matter realization of the chiral anomaly, Nielsen and Ninomiya \cite{Nielsen1983} considered a chiral Weyl linear dispersion of the zero-th Landau levels. Application of an electric field parallel to the magnetic field generates an imbalance between the zero Landau levels of opposite chiralities. This axial current leads to a positive magnetoconductivity \cite{Armitage2017}:
\begin{equation}
\sigma(B)=\sigma+\frac{e^{4}B^{2}}{4\pi^{4}g(E_{F})},
\end{equation}
where $g(E_{F})$ is the density of states at the Fermi level. Remarkably, this expression can be derived both in the quantum limit or in the semi-classical framework without introducing Landau levels \cite{Armitage2017}. We illustrate the chiral anomaly with Weyl fermions in Fig. \ref{fig:11}(a). 

The negative magnetoresistance became accepted as the signature of the presence of linearly dispersing topological quasiparticles, and was possibly observed for instance in GdPtBi \cite{Hirschberger2016}, which is a quadratic gapless semiconductor. In the magnetic field, the fourfold degenerate band-touching splits and pairs of Weyl points are created as we illustrate in schematics in Fig. \ref{fig:11}(b). Rotating the external magnetic field from out-of-plane to in-plane (see Fig. \ref{fig:11}(c)) changes the magnetoresistance from positive to negative. Positive non-saturating magnetoconductance was observed recently also in the Mn$_{3}$Sn AF \cite{Kuroda2017}. Alternative sources of positive magnetoconductance such as current jetting and weak localization were carefully ruled out in this study. The positive magnetoconductance in Mn$_{3}$Sn is linear in a magnetic field which was attributed to the type-II Weyl fermions \cite{Kuroda2017} in contrast to the quadratic magnetoconductance observed in type-I (non-tilted) Weyl semimetals \cite{Zyuzin2017}.  The detailed role of topological quasiparticles in the negative and large non-saturating magnetoresistance remains to be clarified \cite{Ali2014,Pletikosic2014,Soluyanov2015,Khouri2016}. Here the first step was made by \textit{ab initio} \cite{Kim2017e} and transport studies \cite{Zhang2017f} of Weyl points in strong magnetic fields signaling the importance of linear dispersion for the observation of negative magnetoresistance. In strong magnetic fields, the negative magnetoresistance disappears which was attributed to the gapping of Weyl points by the Zeeman splitting.

\subsection{Topological phase transitions and anisotropic magnetoresistance in antiferromagnetic systems}
Topological phase transitions were experimentally demonstrated in heterostructures with TIs and AFs \cite{He2016,He2016a} or systems with an artificially engineered AF coupling \cite{Mogi2017}. AF CrSb/TI (Bi,Sb)$_{\text{2}}$Te$_{\text{3}}$ /AF CrSb heterostructure shows spikes in the magnetoresistance which were attributed to a topological phase transition of Dirac quasiparticles at the interfaces \cite{He2016a}. A MTI/TI/MTI heterostrucutre was reported for a presumed topological phase transition between QAHE state and axion insulator (quantized topological magnetoelectric effect) by switching the magnetic order in the MTI from ferromagnetic to antiferromagnetic by an external magnetic field or electric gating \cite{Mogi2017,Tokura2017}. Albeit at mK temperatures, the phase transition yields very large magnetoresistance or electroresitance changes corresponding to switching on and off the quantized conductivity plateous  $h/e^{2} \sim 25,8 k\Omega$ \cite{Mogi2017,Tokura2017}. The QAHE effect was to date observed only at mK temperatures \cite{Wang2017b}. Searching for novel mechanism and material candidates with a more robust, controllable and room temperature QAHE states represents an important direction of future research in topological spintronics. Here for instance, QAHE induced by electrical gating was predicted in Sr$_{\text{2}}$FeOsO$_{\text{6}}$ AF \cite{Dong2016}.
\begin{figure}[h]
\sidecaption
\includegraphics[width=1\linewidth]{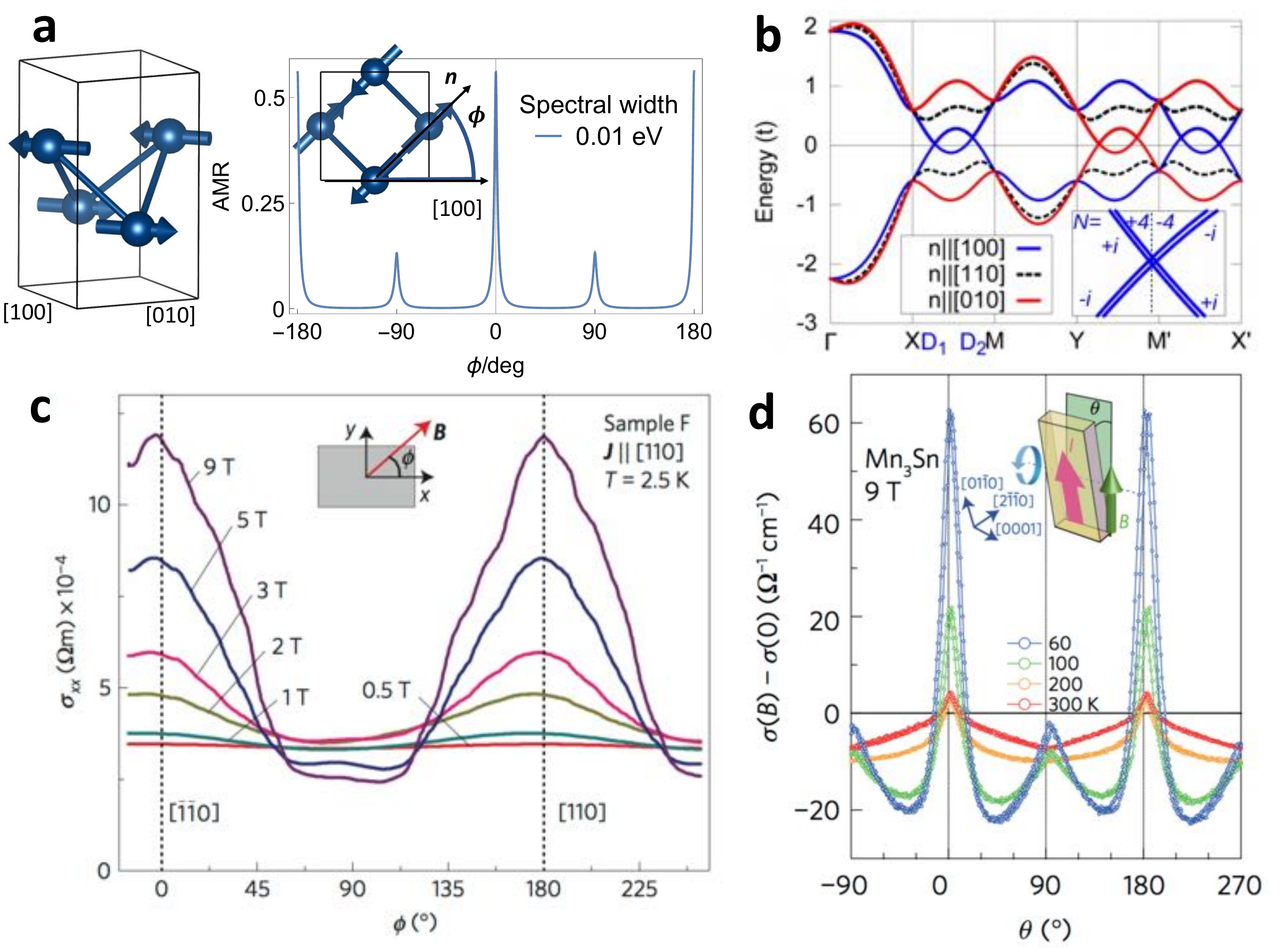}
\caption{\textbf{Extreme anisotropies in magnetoresistance effects in topological semimetals.} (a) Large topological anisotropic magnetoresistance in a tetragonal AF Dirac semimetal model. (b) Band structure of the model Dirac AF. (c) Angular dependence of magnetoresistance in a magnetically induced Weyl semimetal GdPtBi. (d) Angular dependence of magnetoresistance in AF Weyl semimetal Mn$_{3}$Sn. Panel (b) adapted from Ref\cite{Smejkal2016}, panel (c) from \cite{Hirschberger2016}, and panel (d) from \cite{Kuroda2017}.}
\label{fig:12}   
\end{figure}

The prediction of tuning the N\'{e}el order parameter by the N\'{e}el SOT or gating in antiferromagnetic DSMs opens the possibility of using the topological metal-insulator transition (TopoMIT) in a bulk AF \cite{Smejkal2016,Smejkal2017a}. The origin of these effects is in the sensitivity of Dirac crossing hybridations on the orientation of the N\'{e}el vector in the orthorhombic magnetic crystalline symmetry, as shown in Fig. \ref{fig:9}(d). In the presence of SOC, only for the N\'{e}el vector along the [001] axis, protected Dirac fermions emerge. All the other N\'{e}el vector orientations lead to a gapped spectrum at the Fermi level \cite{Smejkal2016}. The transport counterpart of the topoMIT was predicted to be the topological anisotropic magnetoresistance (topoAMR) \cite{Smejkal2016}. The topoAMR can be extremely large and is understood as a limiting case of the crystalline AMR. In Fig. \ref{fig:12}(a) we show the tetragonal lattice of a minimal model of the DSM AF \cite{Smejkal2016}. The corresponding Hamiltonian reads $H_{\textbf{k}}=-2t\tau_{x}\cos\frac{k_{x}}{2}\cos\frac{k_{y}}{2}-t'\left(\cos k_{x}+\cos k_{y}\right) + 
 \lambda \tau_{z}\left( \sigma_{y} \sin k_{x} - \sigma_{x} \sin k_{y} \right) + \tau_{z}J_n\boldsymbol\sigma\cdot\textbf{n}$, where first two terms are first and second neighbor hoppings, third term is a staggered SOC, last term is AF s-d type exchange and $\tau$, and $\sigma$ are Pauli matrices corresponding to orbital and spin-degree of freedom, respectively. The longitudinal conductivity is calculated from the Boltzmann formula in the limit of a small spectral broadening $\Gamma$:
\begin{equation}
\sigma_{xx}(\phi)=\frac{e^{2}}{\hbar 4\Gamma L^{2}}\sum_{\textbf{k}n}\frac{\partial E_{\textbf{k}n}}{\partial k_{x}}\frac{\partial E_{\textbf{k}n}}{\partial k_{x}}\delta(E-E_{F}),
\end{equation}
where $L$ is the size of the system, and  $\phi$ is the angle between [100] axis and magnetization. The AMR is defined as:
\begin{equation}
\text{AMR}=-\frac{\sigma_{xx}(\phi)-\sigma}{\sigma},
\end{equation}
where $\sigma$ is the average conductivity within the plane. In Fig.\ref{fig:12}(a) we show the angular dependence of the AMR in this model \cite{YamamotoSmejkal2017}. In Fig. \ref{fig:12}(b) we show the band structure of the model. For the N\'{e}el vector orientations [100] and [010], preserving the glide mirror planes of the system \cite{Smejkal2016}, the Dirac points are gapless and conduct. Once the N\'{e}el vector is rotated away from these high symmetry axes, the crystalline symmetries are broken, Dirac bands hybridize and gap opens. Consequently the conductivity decreases exponentially. The sharp peaks in the angular dependence are very different when comparing to the standard harmonic AMR dependence in ferromagnetic alloys. Also the origin is very distinct. The topoAMR originates in Fermi surface topology changes instead of scattering effects responsible for the standard AMR. 
The difference in conductivity between the [100] and [010] direction originates in the anisotropy of the Dirac cones. The orthorhombic CuMnAs was predicted as the realistic material candidate based on \textit{ab initio} calculations \cite{Smejkal2016}. The interplay of the Dirac points and topoAMR with disorder, interaction effects, and nonequilibrium currents needs to be carefully addressed to potentially make the effect relevant for real spintronics device applications. Foreseen applications include topological transistors or memories \cite{Smejkal2017a}. %Here the necessity for the both inducing the topological state and controlling the magnetic order is removed. 
Since the control of the N\'{e}el vector can be achieved either by the N\'{e}el SOT due to the applied current or due to the tuning of the MAE by electric gating, the effect is presumably more favorable for spintronics than the MIT manipulated by external magnetic field in pyrochlores \cite{Tian2015} or AF topological semimetal candidate NdSb \cite{Wakeham2016}. Although this topoAMR due to the MIT was not experimentally discovered yet, analogical effects controlled by external magnetic field were observed in WSMs.  

In GdPtBi, Weyl point positions are sensitive to the orientation of the applied magnetic field \cite{Hirschberger2016}. This leads in turn to a pronounced angular dependence of the magnetoconductance, as we show in Fig. \ref{fig:12}(c). 
The changes in magnetoconductance are attributed to the varying angle between the crystalline axis and the Zeeman field what is in contrast to the behavior predicted for the AF DSM CuMnAs. The spikes in magnetoconductance have been measured also in the correlated WSM AF Mn$_{3}$Sn as we illustrate in Fig. \ref{fig:12}(d) \cite{Kuroda2017}. Here the magnetic order is controlled by the relatively weak external magnetic field. The reorientation of the moments changes the local symmetry and possibly redistributes the Weyl points close to the Fermi level \cite{Kuroda2017}. Importantly, the spikes in magnetoconductance persist to temperatures $\sim$100\,K, despite the WSM is highly correlated and disordered. These temperatures are much higher temperatures than the reported QAHE critical temperatures of $\sim$10\,mK. 

\subsection{Anomalous Hall effect in noncollinear antiferromagnets}
AHE refers to the transversal electric current generation in the magnet subjected to an applied longitudinal electric field. The anomalous Hall conductivity is for the magnetization along $z$ axis the antisymmetric part of the conductivity tensor:
\begin{equation}
\sigma_{\text{AHE}}=\frac{\sigma_{xy}-\sigma_{yx}}{2}.
\end{equation}
For a long time, the AHE was considered to scale with the magnetization: $\rho_{H}=R_{0}H_{z}+R_{S}M_{z}$, where the first part corresponds to the ordinary Hall effect due to the external magnetic field $H_{z}$, and the second term is the AHE due to the $\cal{T}$ symmetry breaking due to the magnetization $M_{z}$, and $R_{0}$, and   $R_{S}$ are ordinary and (spontaneous) anomalous Hall coefficients, respectively. AHE was traditionally attributed to the simultaneous presence of $\cal{T}$ symmetry breaking by the ferromagnetism and SOC. Thus, naively, one would expect that the AHE must vanish in AFs because of the compensation moments of the opposite sublattices. Indeed, this picture is valid in simple AFs where the combination of $\cal{T}$ symmetry with another crystalline symmetry forces the AHE to vanish. Typical examples include collinear AFs with $\mathcal{PT}$ or $\mathcal{T}T_{1/2}$ symmetries which we discussed in the context of the Dirac quasiparticles and TIs. 
Remarkably, the AHE was observed in systems with a negligible net magnetization and without the necessity for SOC. 
Interestingly, already Haldane pointed out in 1988 \cite{Haldane1988} the possibility of the quantized AHE in honeycomb lattice with complex hoppings with a staggered potential and Shindou et al. \cite{Shindou2001} later demonstrated the nonzero AHE in a model calculation in AHE induced by distorting the FCC lattice with 3Q AF order. More recently Hua Chen et al. \cite{Chen2014} and K\"{u}bler and Felser \cite{Kubler2014} predicted a large AHE in noncollinear AFs with a negligible net moment. Presumably, the large MAE in Mn$_{3}$Ir AF makes it impossible to orient the magnetic domains and thus prevents the experimental detection of the AHE in this compound. 

\begin{figure}[h]
\sidecaption
\includegraphics[width=1\linewidth]{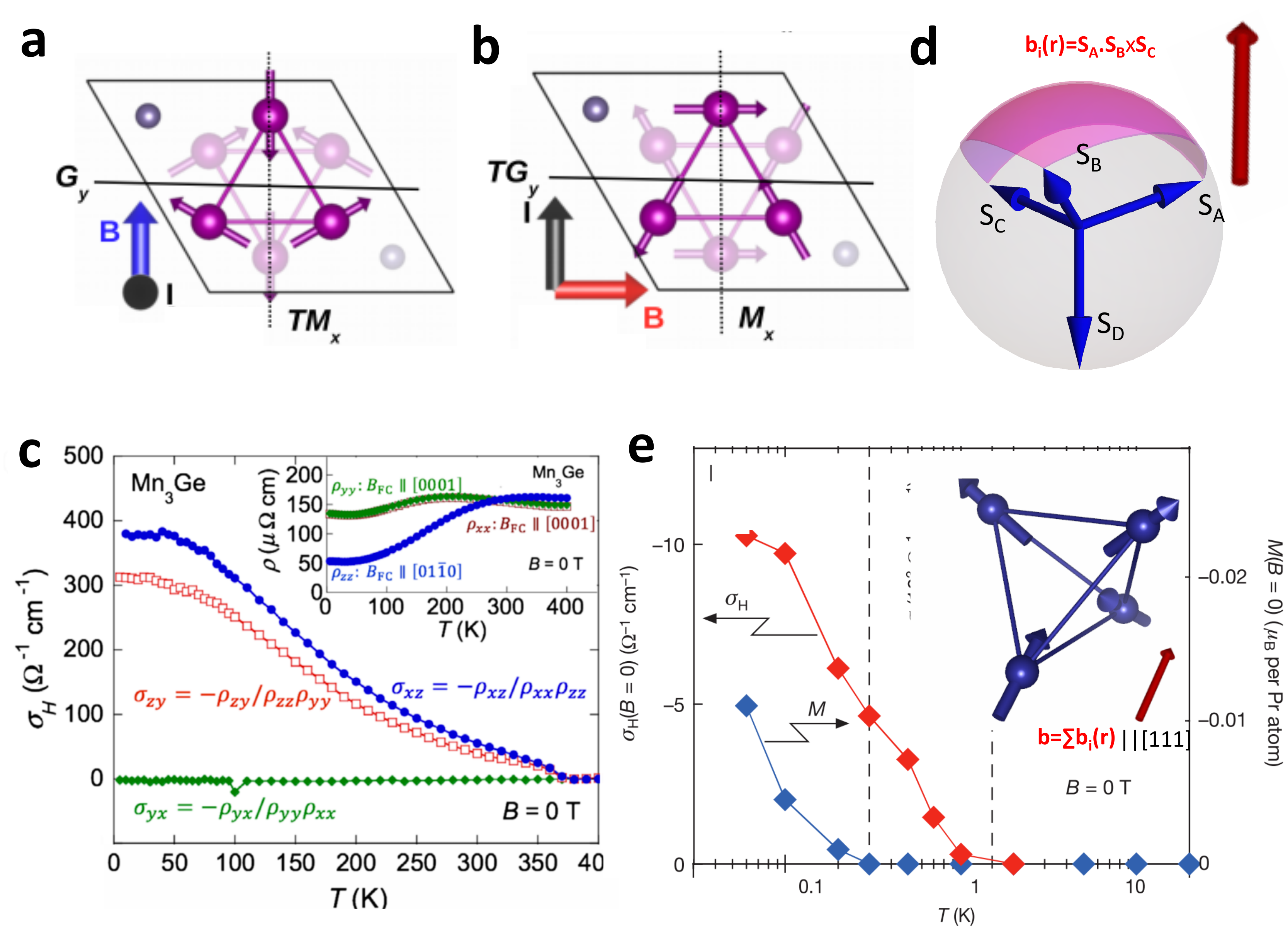}
\caption{\textbf{Anomalous and topological Hall effects in chiral AFs.} Two chiral inverted AF structure stabilized by an external magnetic field along (a) [010], and (b) [100] direction. (c) Measured temperature dependence of AHE in Mn$_{3}$Ge originating from AF texture. (d) Spin chirality and non-coplanar magnetic moments of an antiferromagnetic candidate for the quantum topological Hall effect. (e) Topological Hall effect observed in the spin liquid state on a pyrochlore lattice - fragment shown in the inset. Panels (a,b) adapted from \cite{Smejkal2017}, panel (c) from \cite{Kiyohara2015}, and panel (e) from \cite{Machida2010}.}
\label{fig:13}       
\end{figure}

However, Mn$_{3}$Ge and Mn$_{3}$Sn have a much smaller MAE. AHE from the AF texture, $\rho_{H}^{AF}$, was indeed observed in these compounds by carefully substracting the Hall effect originating from the external field, $R_{0}H_{z}$, and from the small ferromagnetic moment, $R_{S}M$ \cite{Nakatsuji2015,Nayak2016}:
\begin{equation}
\rho_{H}=R_{0}H_{z}+R_{S}M+\rho_{H}^{AF},
\label{hall_exp_af}
\end{equation} 
where $M$ is the net magnetization.  The experimental value of the AHE in Mn$_{3}$Ge is $\sigma_{xz}$  $\approx$ 380\,$\Omega^{-1}\text{cm}^{-1}$ \cite{Kiyohara2015} while the \textit{ab initio} calculation from Berry curvature gives $\sigma_{xz}\approx$ 330\,$\Omega^{-1}\text{cm}^{-1}$ \cite{Zhang2016d}. 
The noncollinear AF order on the kagome lattice breaks the time-reversal symmetry as we show in Fig. \ref{fig:13}(a) and \ref{fig:13}(b). For the AF structure in Fig.~\ref{fig:13}(a)  \cite{Tomiyoshi1982,Kiyohara2015} there is an effective time reversal symmetry $\mathcal{T}\mathcal{M}_{x}$ ($\mathcal{M}_{x}$ is the mirror (100) plane symmetry) which gives $\sigma_{yz}=0$, and the emergent magnetic field lies along $\textbf{B}\parallel$[010], and only $\sigma_{xz}\neq 0$. In constrast, for the chiral texture in Fig.~\ref{fig:13}~(b) \cite{Tomiyoshi1982,Kiyohara2015}, the effective $\mathcal{T}\mathcal{G}_{y}$ symmetry ($\mathcal{G}_{y}$ is the glide mirror plane $\left\lbrace \mathcal{M}_{y} \vert (0,0,\frac{1}{2}) \right\rbrace$) implies $\sigma_{xz}=0$, the emergent magnetic field points along $\textbf{B}\parallel$[100], and only $\sigma_{zy}\neq 0$. Furthermore, independent on the in-plane orientation, there is an effective time reversal symmetry $ \mathcal{T}\mathcal{M}_{z}$ ($\mathcal{M}_{z}$ is the mirror (001) plane symmetry) making the component
$\sigma_{xy}=0$. The symmetry analysis is consistent with the experimental data measured on Mn$_{3}$Sn and Mn$_{3}$Ge and presented in Fig. \ref{fig:2}(b) and \ref{fig:13}(c). In conclusion, the spin-orbit entangled bands generate a large fictitious magnetic field in the crystal momentum space parallel to the direction of the field stabilizing the triangular order, and the AHE takes place in the plane perpendicular to the field.   

The emergent magnetic field was estimated to be very large, of the order of $\sim$ 200\,T in Mn$_{3}$Ge \cite{Kiyohara2015}. Although the Mn$_{3}$Ge and Mn$_{3}$Sn were predicted to host Weyl points close to the Fermi level, the \textit{ab initio} calculation of the AHE shows that the largest contributions come from BZ regions not related to any identified Weyl points, but rather from spin-orbit entangled avoided crossings \cite{Zhang2016d}. A recent study by K\"{u}bler and Felser \cite{Felser2017}, however, demonstrates the possibility of propagation of Fermi sea Weyl points in Mn$_{3}$Ge and Mn$_{3}$Sn to the Fermi surface-quasiparticle transport \cite{Haldane2004,Gos2015}. 

\subsection{Topological Hall effects and antiferromagnetic skyrmions}
In the topological Hall effect, the role of SOC is overtaken by the spin chirality. We show in Fig.\ref{fig:13}(d) the spin chirality generating a nonzero Berry curvature. Spin chirality is nonzero in non-coplanar spins, in contrast, it vanishes in coplanar non-collinear antiferromagnetic structures of, e.g., Mn$_{3}$Ge and Mn$_{3}$Sn. The spin chirality generates a fictitious magnetic field (see red arrow in Fig. \ref{fig:13}(d)), $\hat{\textbf{m}}\cdot \left(\partial_{x}\hat{\textbf{m}} \times \partial_{y}\hat{\textbf{m}}\right)$. This field acting on the Bloch electrons generates a Hall response. The topological Hall effect and the AHE can be possible to experimentally disentangled by analyzing the disorder dependence \cite{Kanazawa2011}. The topological  Hall effect was initially reported in antiferromagnetic pyrochlore iridates (see Fig.\ref{fig:13}(e))
 \cite{Machida2010} and later in MnSi chiral antiferromagnetic alloys \cite{Surgers2014,Surgers2016}. We note that the effect does not imply in this case a correspondence to a topological invariant, in sharp contrast to the topological Hall effect in skyrmions. However, in the quantum topological Hall effect proposed for the non-coplanar AF K$_{0.5}$RhO$_{2}$ \cite{Zhou2016}, with its magnetic sublattices shown schematically in Fig.\ref{fig:13}(d), the topological charge occurs in the momentum space as in the QAHE. 

The topological Hall effect from a skyrmion spin texture is associated with a topological winding number of the skyrmion \cite{Tokura2017}. It is important to distinguish it from the skyrmion Hall effect which refers to the deflection of skyrmion center due to the Magnus force. The Magnus force is according to micromagnetic calculations not present in AF skyrmions, which implies that AF skyrmions might move in straight lines \cite{Barker2016}. This is favorable when considering skyrmions for storing information in racetracks. 
Finally, the topological Hall effect in AF skyrmions with a compensating sublattices vanishes, while the topological spin Hall effect can be still sizable and can be used to detect an AF skyrmion \cite{Buhl2017,Gobel2017}. 

\section{Summary}
Antiferromagnetic spintronics has been recently established as a new branch of magnetism \cite{MacDonald2011,Jungwirth2016,Smejkal2017,Sander2017a}. In parallel, last few years have seen progress in coupling magnetism with topological states of matter, giving rise to a new spin-off: topological spintronics \cite{Fan2016b}. We have shown that AF order might play an important role in topological spintronics due to the unique AF symmetries \cite{Smejkal2017a}. While signatures of correlated AF WSM were already observed \cite{Kuroda2017}, other topological AF phases remain to be discovered. The large signal to noise ratio was reported in magnetoresistance signals of topological semimetals \cite{Smejkal2016,Kuroda2017}. Further theoretical and experimental progress will possibly lead to topological spintronics effects improving the reading and writing signals in AFs \cite{Smejkal2017a}. The progress in writing efficiency due to the nontrivial topologies is in its infancy, although an increase of the spin Hall angle in TI/FM\cite{Mellnik2014}, TI/MTI\cite{Fan2014a}, or TI/ferrimagnetic systems \cite{Han2017} was already reported and novel mechanisms for dissipationless SOT were suggested \cite{Hanke2017a}. Here we focused on the state-of-the-art effects which were predicted, and some of them already experimentally confirmed, in antiferromagnetic systems. The unique AF symmetries and the abundancy of AF allow for other research directions to emerge such as topological superconductivity in AFs and spintronics based on antiferromagnetic skyrmions \cite{Smejkal2017a}.

\begin{acknowledgement}
We acknowledge support from the Ministry of Education of the Czech Republic Grants LM2015087 and LNSM-LNSpin, the Grant Agency of the Czech Republic Grant No. 14-37427,  and the EU FET Open RIA Grant No. 766566.
\end{acknowledgement}

\bibliographystyle{spphys}

\end{document}